\documentclass[a4paper,fleqn,usenatbib]{mnras}

\usepackage{graphicx}
\usepackage[percent]{overpic}
\usepackage{lipsum}
\usepackage{multirow}
\usepackage{color}
\usepackage{rotating}
\usepackage{mwe,tikz}
\usepackage[percent]{overpic}
\usepackage{mathtools}
\usepackage{physics}
\usepackage{amsmath}
\usepackage[utf8x]{inputenc}
\usepackage{hyperref}
\usepackage{wasysym}
\hypersetup{
    colorlinks=true,
    linkcolor=blue,
    filecolor=magenta,      
    urlcolor=blue,
}
\usepackage{soul}

\def\ltsima{$\; \buildrel < \over \sim \;$}
\def\simlt{\lower.5ex\hbox{\ltsima}}
\def\gtsima{$\; \buildrel > \over \sim \;$}
\def\simgt{\lower.5ex\hbox{\gtsima}}
\def\kms{{\rm\,km\,s^{-1}}}
\def\pc{{\rm\,pc}}
\def\kpc{{\rm\,kpc}}

\newcommand{\fmmm}[1]{\mbox{$#1$}}

\newcommand{\mcnd}{\mbox{\fmmm{'}\hskip-0.3em .}}
\newcommand{\mcnp}{\mbox{\fmmm{'}}}

\def\deg{^\circ}
\def\degg{\hbox{$\null^\circ$\hskip-3pt .}}

\def\Gyr{{\rm\,Gyr}}

\def\masyr{{\rm\,mas \, yr^{-1}}}

\def\ltsima{$\; \buildrel < \over \sim \;$}
\def\gtsima{$\; \buildrel > \over \sim \;$}

\title[Stellar Streams in Gaia DR2]{Ghostly Tributaries to the Milky Way: Charting the Halo's Stellar Streams with the Gaia DR2 catalogue}

\author[Malhan, Ibata \& Martin]{
Khyati Malhan,$^{1}$\thanks{E-mail: khyati.malhan@astro.unistra.fr}
Rodrigo A. Ibata,$^{1}$\thanks{E-mail: rodrigo.ibata@astro.unistra.fr}
Nicolas F. Martin,$^{1,2}$\thanks{E-mail: nicolas.martin@astro.unistra.fr}
\\
$^{1}$Universit\'e de Strasbourg, CNRS, Observatoire Astronomique de Strasbourg, UMR 7550, F-67000 Strasbourg, France\\
$^{2}$Max-Planck-Institut f\"ur Astronomie, K\"onigstuhl 17, D-69117 Heidelberg, Germany\\
}

\date{Accepted 2018 September 05. Received 2018 August 11; in original form 2018 May 01}

\pubyear{2018}

\begin{document}
\label{firstpage}
\pagerange{\pageref{firstpage}--\pageref{lastpage}}
\maketitle

\begin{abstract}
We present a panoramic map of the stellar streams of the Milky Way based upon astrometric and photometric measurements from the Gaia DR2 catalogue. In this first contribution, we concentrate on the halo at heliocentric distances beyond $5\kpc$, and at Galactic latitudes $|b|>30\deg$, using the \texttt{STREAMFINDER} algorithm to detect structures along plausible orbits that are consistent with the Gaia proper motion measurements. We find a rich network of criss-crossing streams in the halo. Some of these structures were previously-known, several are new discoveries, but others are potentially artefacts of the Gaia scanning law and will require confirmation. With these initial discoveries, we are starting to unravel the complex formation of the halo of our Galaxy.
\end{abstract}


\section{Introduction}

The central position that stellar streams hold for Galactic Archeology studies motivates conducting a thorough census of such structures in the Milky Way. Besides testing the hierarchical merging scenario of Galaxy formation \citep{Johnston1996, Helmi1999BuildingHalo}, the number of stellar streams can, in principle, be used to put a lower limit on past accretion events into the Galactic halo, their orbital structures can be used to  probe the mass distribution and shape of the Milky Way dark matter halo \citep{Johnston1999GalPot, Ibata2001,  EyreBinney2009, Koposov2010,LawMajewski2010, Kupper2015, Bovy2016GD1Pal5}, stream-gaps can provide indirect evidence for the existence of dark matter sub-halos \citep{Johnston2002subhalos, StreamGap_Carlberg2012, StreamGap_Erkal2016, StreamGap_Sanders2016}, and these structures can also be used to constrain the models of the formation and evolution of globular clusters \citep{Balbinot2018M}. Furthermore, analyses based on the quantity and the collective phase-space distribution of stellar streams hold great promise in addressing some small-scale $\Lambda$CDM problems (such as the ``missing satellite problem'' and the ``plane-of-satellites'' problem, see, e.g. \citealt{Bullock2017problems}).

\begin{figure*}
\begin{center}
\includegraphics[width=\hsize]{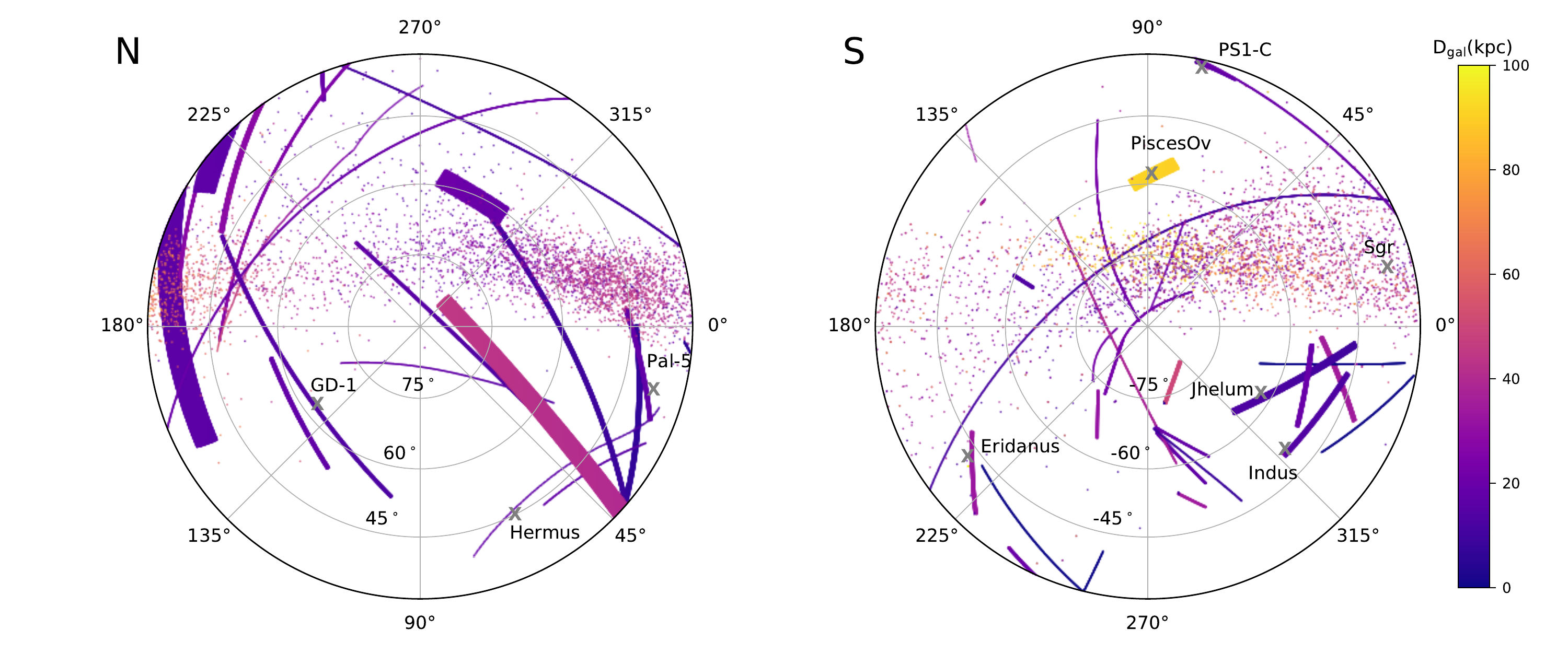}
\vspace{-1.0cm}
\end{center}
\caption{Schematic stellar Stream map of the Milky Way sky prior to Gaia DR2. Here we show the Milky Way stellar stream map (minus some stellar clouds) from the \texttt{GALSTREAMS} package \citep{Mateu2017}, transformed into polar ZEA projections. The colour represents the Galacto-centric distances to these structures. The left and right panels show, respectively, the projection from the North and South Galactic poles. The names of a few streams are labelled to help the reader's orientation in this coordinate system. Galactic longitude increases clockwise in the north panel and counter-clockwise in the south panel, while Galactic latitude changes radially as shown.}
\label{fig:ZEA_Cecilia}
\end{figure*}

Such considerations have motivated many previous studies to detect and analyse stellar streams in our Galaxy. Notable efforts in the past include the ``Field-of-streams'' map \citep{Belokurov2006} of the region around the North Galactic Cap based on the SDSS DR5, which was expanded to cover both the Northern and Southern Galactic Cap regions in later SDSS releases (see, e.g. \citealt{Grillmair2016BookChapter}); \cite{Bernard2014} created a panoramic map of the entire Milky Way halo north of $\delta \sim -30 \deg$ ($\sim 30,000\,\rm{deg^2}$) based on the Pan-STARRS1 dataset; \cite{Mateu2017} applied a pole-counts stream-finding method to the Catalina RR-Lyrae survey revealing 14 candidate streams in the inner Galaxy; and most recently \cite{DES_11_streams2018} discovered 11 stellar streams out to a distance of $d_{\odot}\sim 50\kpc$ by making use of the data from the Dark Energy survey (DES). The regions of sky covered by presently-known streams have been conveniently compiled in the \texttt{GALSTREAMS} python package \citep{Mateu2017}, which we reproduce in Figure \ref{fig:ZEA_Cecilia} for comparison to our results.

Given the arrival of all-sky data of unprecedented astrometric quality from the ESA/Gaia survey \citep{Gaia2012deBruijne, GaiaDR12016}, we built a stream-finding algorithm (the   \texttt{STREAMFINDER}, \citealt{Malhan2018_SFI}, hereafter Paper~I) to make use of the kinematic information that Gaia provides. The idea that we incorporated in the \texttt{STREAMFINDER} algorithm is that stellar streams can be found more efficiently by searching along possible orbital trajectories in the underlying gravitational potential of the Galaxy. In Paper I, our tests, based on a suite of N-body simulations embedded in a mock Galactic survey, showed that the algorithm is able to detect distant halo stream structures containing as few as $\sim 15$ members (or equivalently with a surface brightness as low as $\Sigma_{\rm G} \sim 33.6\, {\rm mag \, arcsec^{-2}}$ ) in the End-of-mission Gaia dataset. The detection limit depends on various factors, such as the stream structure itself and its location in phase-space with respect to the contaminating background. For instance, in \cite{Ibata_2018_Phlegethon} we reported the discovery of the (high contrast) Phlegethon stream in Gaia DR2 with a surface brightness of $\Sigma_G\sim 34.6 \, {\rm mag \, arcsec^{-2}}$.

The purpose of this contribution is to present an updated stellar stream map of the halo of the Milky Way (at $D_\odot>5\kpc$) obtained via the application of our \texttt{STREAMFINDER} algorithm onto the recently published Gaia Data Release 2 (DR2) 
\citep{GaiaDR2_2018_Brown,GaiaDR2_2018_astrometry,GaiaDR2_2018_Parallaxes,GaiaDR2_2018_photometry,GaiaDR2_2018_Pmotions}. In this first analysis, we restrict ourselves to analysing the outer halo at distances beyond $5\kpc$ as our algorithm takes longer to compute in inner regions where the density of both the field stars and the possible candidates is large (as is the case when considering closer structures or indeed in the vicinity of the Galactic Plane). 

The paper is organized as follows: Section~\ref{sec:Data_sample} details the selections made on the Gaia data; Section~\ref{sec:contamination_model} explains how we built a model of the contaminating populations of the Milky Way; the analysis using our \texttt{STREAMFINDER} algorithm is detailed in Section~\ref{sec:streamfinder_analysis}; the results are presented in Section~\ref{sec:Results}; finally we discuss these findings and draw our conclusions in Section~\ref{sec:conclusions}.

\begin{figure*}
\begin{center}
\includegraphics[viewport= 135 40 692 320, clip,width=0.8\hsize]{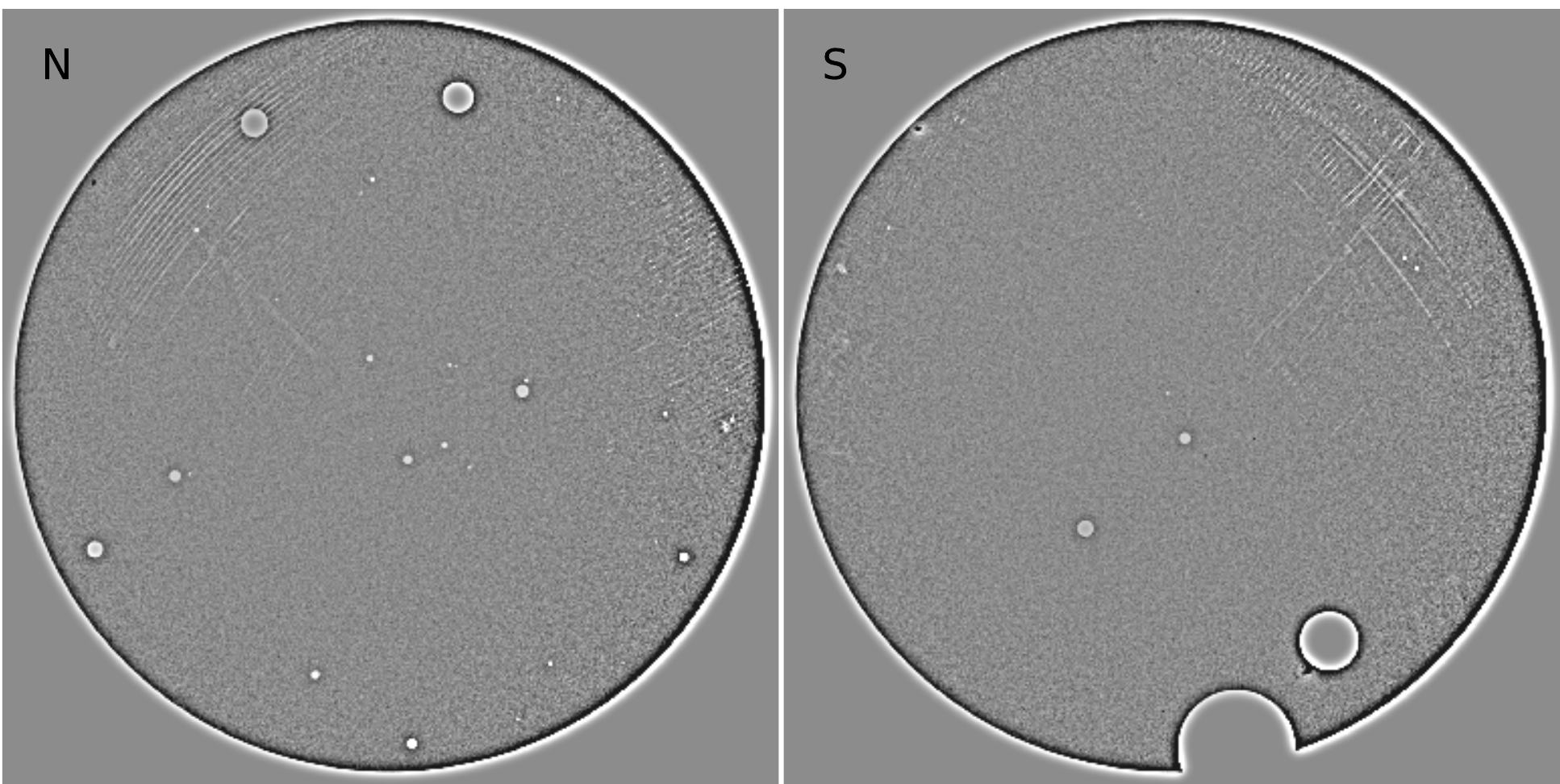}
\end{center}
\caption{Unsharp-mask map of the $|b|>10\deg$ sky, derived using Gaia sources brighter than ${\rm G_0=20}$. This simple filtering procedure highlights the stripy artefacts that arise due to the inhomogeneous scanning of the sky in the DR2 catalogue. The same ZEA projections are used here as in Figure~\ref{fig:ZEA_Cecilia}. To create this map, we binned the catalogue into pixels of size $5\mcnd3\times5\mcnd3$, and subtracted from this the same map but smoothed with a two-dimensional Gaussian of standard deviation $53\mcnp$. The holes seen in the image correspond to the excised regions around known clusters and satellite galaxies that were omitted in our analysis.}
\label{fig:Gaia_scanning}
\end{figure*}

\section{Data and Stream Search Analysis}\label{sec:Data_sample}

We use the Gaia DR2 catalogue for all of our present analysis. This dataset provides positions, parallaxes and proper motions (a 5D astrometric solution) for over 1.3 billion stars down to ${\rm G\sim20.7}$ in our Galaxy, along with the Gaia broad-band photometry in the ${\rm G, G_{BP}, G_{RP}}$ pass-bands. The information that is useful for our purpose are the stellar positions ($\alpha, \delta$), parallaxes ($\pi$), proper motions ($\mu_{\alpha}, \mu_{\delta}$), magnitudes (${\rm G, G_{BP}, G_{RP}}$) and the associated observational uncertainties. 

We correct all Gaia sources from extinction using the \cite{Schlegel_Extinc_maps1998} maps, assuming $A_{\rm G}/A_{\rm V}=0.85926$, $A_{\rm BP}/A_{\rm V}=1.06794$, $A_{\rm RP}/A_{\rm V}=0.65199$\footnote{These extinction ratios are listed on the Padova model site \url{http://stev.oapd.inaf.it/cgi-bin/cmd_2.8}.}. Doing so, we naturally assume that the extinction is entirely in the foreground of the studied stars, which is likely a good assumption for the halos stars we analyse here. Henceforth, all magnitudes will refer to the extinction-corrected values.

The Gaia DR2 is based on only 22 months of observations, and not all areas of sky have been observed to uniform depth. Gaia scans the sky while spinning, and this naturally imprints great circles into the depth map. In Figure~\ref{fig:Gaia_scanning} we show the result of applying an unsharp-mask to all data at Galactic latitudes $|b|>10\deg$ and with ${\rm G<20}$. A large number of stripy residuals can be seen, which could in principle masquerade as streams. Any structures following this pattern are almost certainly artefacts.

After extensive tests of the \texttt{STREAMFINDER} using the Gaia Universe Model Snapshot (GUMS, \citealt{GUMS2012}), we decided to limit the sample for the present contribution to $|b|>30\deg$ and $G<19.5$. The chosen magnitude limit mitigates against the effect of completeness variations due to inhomogenous extinction, while also reducing the number of sources that need to be examined. Likewise, the Galactic latitude constraint also greatly diminishes the size of the sample. We retained only those  sources that had a full 5-parameter astrometric solution, along with valid magnitudes in all three photometric bands.

We further omitted all Gaia DR2 catalogue stars within two tidal radii of the Galactic globular clusters listed in the compilation by \cite{Harris2010_MW_GC}, as well as all the stars within 7 half-light radii around Galactic dwarf satellite galaxies (as compiled by \citealt{McConnachie2012}). This was implemented so as to avoid creating spurious stream detections that might be caused by the presence of a compact over-density of stars in a given region of phase-space rather than an actual extended stream of stars.

As described in Paper I, it is convenient to reject disk contaminants based on parallax information since we are interested in finding halo structures. The number of these potential nearby contaminants was reduced by removing those sources whose parallax is greater than $1/3000 \, {\arcsec}$ at more than the $3\sigma$ level (i.e. objects that are likely to be closer than $3\kpc$). 

We feed this filtered data to the \texttt{STREAMFINDER}.

\section{Contamination Model}\label{sec:contamination_model}

Before running the \texttt{STREAMFINDER}, we first calculate an empirical smooth model of the Milky Way ``contamination'' (i.e. a model of the smoothly-varying population of stars that lie both in the foreground and the background of the stream-like structures of interest). This contamination model is used as a global probability density function estimate to calculate the likelihood function for identifying substructures. The procedure will be more fully explained in a future contribution (Ibata et al. 2018, in prep.), but briefly, we construct a library of number-density maps of the Galaxy as a function of ${\rm G_{BP}-G_{RP}}$ colour and ${\rm G}$ magnitude in polar Zenithal Equal Area projection with a pixel scale of $1\degg4\times1\degg4$, which are smoothed on a spatial scale of $2\deg$. Furthermore, over spatial regions of $5\degg6\times5\degg6$ (also in polar ZEA projection), we fit the four-dimensional distribution of ${\rm G_{BP}-G_{RP}}$ colour, ${\rm G}$ magnitude, and proper motion $\mu_\alpha$, $\mu_\delta$, with a Gaussian Mixture Model (GMM), with 100 Gaussian components, using the \texttt{Armadillo} C++ library \citep{gmm2017armadillo}. Together, the density maps and the GMM fitted maps allow one to estimate the smoothed probability of finding a star in the Galaxy in the 6-D parameter space of $\alpha, \delta, {\rm G_{BP}-G_{RP}}, {\rm G}, \mu_\alpha, \mu_\delta$.

\section{\texttt{STREAMFINDER} Analysis}\label{sec:streamfinder_analysis}

The \texttt{STREAMFINDER} algorithm is built to detect dynamically cold and narrow tidal stellar streams that are possible remnants of globular clusters or very low-mass galaxies. At the position of every star in the dataset, the algorithm finds the most likely stream model given the observed phase-space information, and quantifies the likelihood of that stream model given the pre-calculated contamination model. To build the stream model, the algorithm launches orbits from the sky position of the star in question, sampling over the proper motion uncertainties, and over the full range of radial velocity. All possible distances to the star are examined that are consistent with the observed photometry and the chosen stellar populations template. A by-product of the algorithm is the orbital solution of every star along which stream lies (see Paper~I for detailed discussion on the workings of the algorithm).

We used the \texttt{STREAMFINDER} to analyse the Gaia DR2 data in a similar way to that described in Paper~I. The orbits are integrated within the Galactic potential model 1 of \cite{Dehnen1998Massmodel}, and these orbits are then projected into the heliocentric frame of observables for comparison with the data. For this coordinate transformation, we assume a Galactocentric distance of the Sun of $8.20\kpc$ \citep{R_sun_value}, a circular velocity $V_{\rm circ} = 240\kms$ and in addition we adopt the Sun's peculiar velocity to be $\bmath{V_{\odot}} = (u_{\odot}, v_{\odot}, w_{\odot}) = (9.0, 15.2, 7.0) \kms$ \citep{Reid_Sun_circ_vel, Schornich2010_Sun}. As explained in Paper I, \texttt{STREAMFINDER} uses pre-selected isochrone models in order to sample orbits in distance space. The selected isochrone model(s) essentially correspond to the proposed Single Stellar Population (SSP) model of the stream.  Here, we chose to work with Padova SSP models \citep{Marigo2008Padova} in the Gaia photometric system, with age $10\Gyr$ and with 7 metallicity values between ${\rm [Fe/H]}=-2.2$ to ${\rm [Fe/H]}=-1.0$ (spaced at $0.2$~dex intervals). These isochrone models cover plausible values for Milky Way halo globular clusters (from which stellar streams are ultimately derived)\footnote{In subsequent papers, we plan to run the algorithm over a fine grid in metallicity and age.}. The candidate model streams were selected to have a Gaussian width of $100\pc$, and to be $10\deg$ long on the sky. Other parameter ranges used to integrate orbits in the Galaxy were identical to those detailed in Paper I.

In Paper~1, our analysis was restricted to a small and relatively high latitude patch of sky ($\sim100\,\rm{deg^2}$) in which the background stellar distribution (the halo) could be approximated as a uniform distribution. In the present case, where we are analysing vast regions of sky that have a non-uniform stellar distribution, it is important to consider the background model of the Galaxy. Therefore, in contrast to Paper~I, the likelihood function that we use here takes the Galaxy into consideration via the smooth contamination model discussed above. Our log-likelihood function is simply:
\begin{equation}
{\ln} \mathcal{L} = \rm \sum_{\rm{data}} {\ln} \, (\eta \mathcal{P}_{\rm signal}  (\theta) + (1 - \eta) \, \mathcal{P}_{\rm contamination} ) \, ,
\end{equation}
where $\theta$ are the stream fitting parameters, $\mathcal{P}_{\rm contamination}$ is the probability density function of the smooth contamination model that we obtain as explained in Section \ref{sec:contamination_model}, and $\eta$ is the fraction of the stream model compared to the contamination. The adopted stream probability density function $\mathcal{P}_{\rm signal}$ is extremely simple: we take the trial orbit under consideration and make it fuzzy by convolving it with a Gaussian in each observed dimension. The Gaussian dispersions are: $\sigma_{\rm sky}$ representing the thickness of the stream, $\sigma_{\mu_{\alpha}}, \sigma_{\mu_{\delta}}$ representing the dispersions in proper motion, and $\sigma_{DM}$ representing the dispersion in distance modulus (and hence in photometry). All these dispersions are the convolution of the intrinsic Gaussian dispersion of the stream model together with the observational uncertainty on each star in the Gaia DR2 catalogue.

\begin{figure*}
\begin{center}
\includegraphics[width=\hsize]{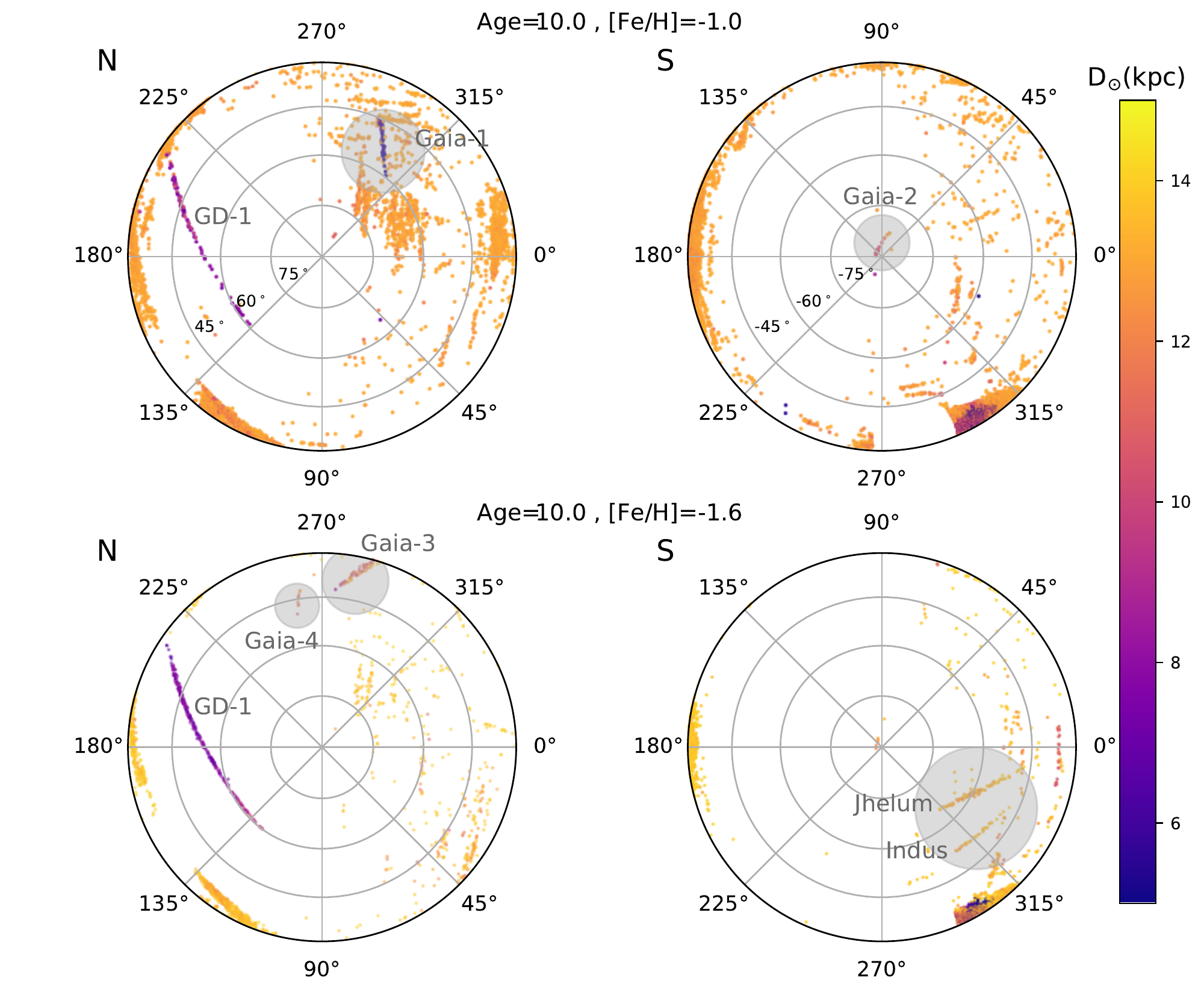}
\vspace{-1.0cm}
\end{center}
\caption{Potential stream stars identified by \texttt{STREAMFINDER} in the inner halo, from $5$ to $15\kpc$, in the same projection as Figure~\ref{fig:ZEA_Cecilia}. The colour represents the distance solutions that are obtained as a by-product for these stars from the \texttt{STREAMFNIDER} analysis. The top panels show a metal-rich selection, while the lower panels show the results for intermediate metallicity.  The most striking structure detected in this distance range is the GD-1 stream \citep{Grillmair2006GD1_correct}, seen clearly towards the lower end of the distance range (coloured purple) in the Northern hemisphere (left panels). Several other streams are visible, including the Jhelum and Indus streams discovered in the DES \citep{DES_11_streams2018}. All stream points displayed here have detection significance $>5\sigma$. New high confidence stream detections are marked on the map, while the others will require confirmation with radial velocity measurements.}
\label{fig:ZEA_inner_halo}
\end{figure*}

\begin{figure*}
\begin{center}
\includegraphics[width=1.0\hsize]{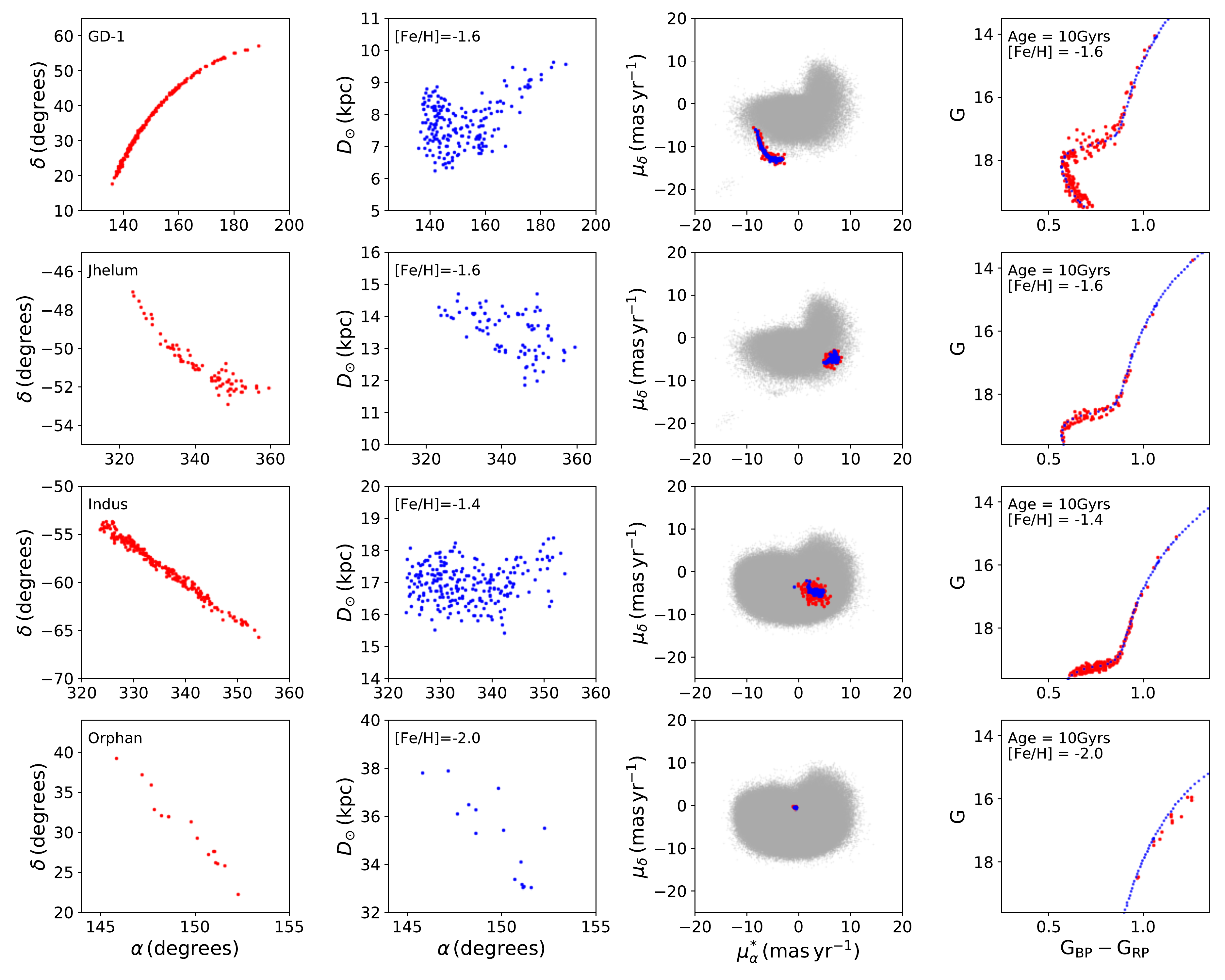}
\end{center}
\caption{Properties of a sample of previously-discovered streams, as recovered by the \texttt{STREAMFINDER}. The first, second, third and fourth rows show the properties of the GD-1, Jhelum, Indus and Orphan streams, respectively. The columns reproduce, from left to right, the equatorial coordinates of the structures, the distance solutions found by the algorithm (for representative metallicity values), the proper motion distribution (with observations in red, model solutions in blue, and the full DR2 sample in grey), and the colour-magnitude distribution of the stars (with observations in red and template model in blue) selected by \texttt{STREAMFINDER}. The distance solutions found by the algorithm match closely the distance values that have been previously derived for these streams: $D_{\odot}\sim8\kpc$ for GD-1 \citep{Grillmair2006GD1_correct}, $D_{\odot}\sim13.2\kpc$ and $\sim16.6\kpc$ for Jhelum and Indus, respectively \citep{DES_11_streams2018} and $D_{\odot}=[33-38]\kpc$ for Orphan \citep{Newberg2010OrphanFit}. The CMD template models, shown in blue in the last column, have been plotted at the appropriate distance for the respective streams. The colour-magnitude diagram of the Orphan stream might seem peculiar, but here we only see the red-giant branch due to the trimming of the data sample below ${\rm G=19.5}$.}
\label{fig:Known_streams}
\end{figure*}

\section{Results}\label{sec:Results}

In Figure \ref{fig:ZEA_inner_halo} we show, for two representative metallicity values, the spatial distribution of the stars in the processed sample that have a high-likelihood of belonging to a stream structure. These data are selected as having ${\ln} \mathcal{L}_{\rm max}/{\ln} \mathcal{L}_{\eta=0} > 15$, where $\mathcal{L}_{\eta=0}$ is the model likelihood when no stream is present, and $\mathcal{L}_{\rm max}$ is the maximum likelihood stream solution found by the algorithm. Thus, our criterion corresponds to $>5\sigma$ when the noise distributions are Gaussian. We would like to point out that the ${\ln} \mathcal{L}_{\rm max}/{\ln} \mathcal{L}_{\eta=0}$ likelihood ratio is calculated for every star in the (filtered) catalogue, yet in the maps presented here we only show those stars where this value exceeds $15$. Many other neighbouring stars may partake in a given stream structure, contributing to the high ${\ln} \mathcal{L}_{\rm max}/{\ln} \mathcal{L}_{\eta=0}$ valued-points marked in the figure, yet they may not themselves pass the criterion and so are not shown. A given stream-like structure seen in the figure is thus composed of many $>5\sigma$ points. However, the points are not statistically independent, as by construction information is correlated over the chosen $10\deg$ trial stream length. We further stress that the aim of the \texttt{STREAMFINDER} algorithm is to enable the detection of streams; a complete characterization and statistical analysis of a given detection should be accomplished with other tools, for instance, by careful modelling with N-body simulations.

The left and right panels of Figure \ref{fig:ZEA_inner_halo} show, respectively, the projection from the North and South poles. The distance solutions displayed here are the ones obtained by the algorithm and span the inner halo range $D_{\odot}= [5,15]\kpc$. The most visible feature in the northern hemisphere is the GD-1 stellar stream \citep{Grillmair2006GD1_correct,Boer2018}, which appears as a $>60\deg$ stream in these spatial density maps of candidate stream members. It is possible that it continues to lower Galactic latitude, where we have not yet run the algorithm. Other notable detections are the Jhelum and Indus streams \citep{DES_11_streams2018} seen in the Southern hemisphere in the more metal-poor map. As a demonstration of the power of the algorithm, we display the properties of GD-1, Jhelum and Indus, as recovered by the \texttt{STREAMFINDER}, in  Figure \ref{fig:Known_streams}. Note that the distance solutions to these streams that we obtain from \texttt{STREAMFINDER} match closely the distance values that have been previously derived for these streams (as explained in Figure~\ref{fig:Known_streams}). The scatter in the distance solutions that is notably seen for individual streams could be a combination of the true intrinsic dispersion of the stream and errors from mismatches with the isochrone template model (from which the distance solutions are derived, see Paper I). We summarize some of the properties of these structures in Table \ref{tab:Stream_properties}, providing, for the first time, the proper motion values for the Jhelum and Indus streams.

The recovery of known stellar streams provides validation of our algorithm. Many other stream-like features can also be seen in this map, but these structures require detailed kinematic analysis for their confirmation (which is beyond of the scope of this paper). In the present contribution we will discuss the most obvious stream structures that not only have coherent phase-space properties (consistent with the template model and the data uncertainties) but that also stand out significantly from the background. These new streams, that are named Gaia-1,2,3,4, are shaded in grey in Figure \ref{fig:ZEA_inner_halo} and their phase-space properties are presented in Figure \ref{fig:New_halo_candidates}.

\begin{figure*}
\begin{center}
\includegraphics[width=\hsize]{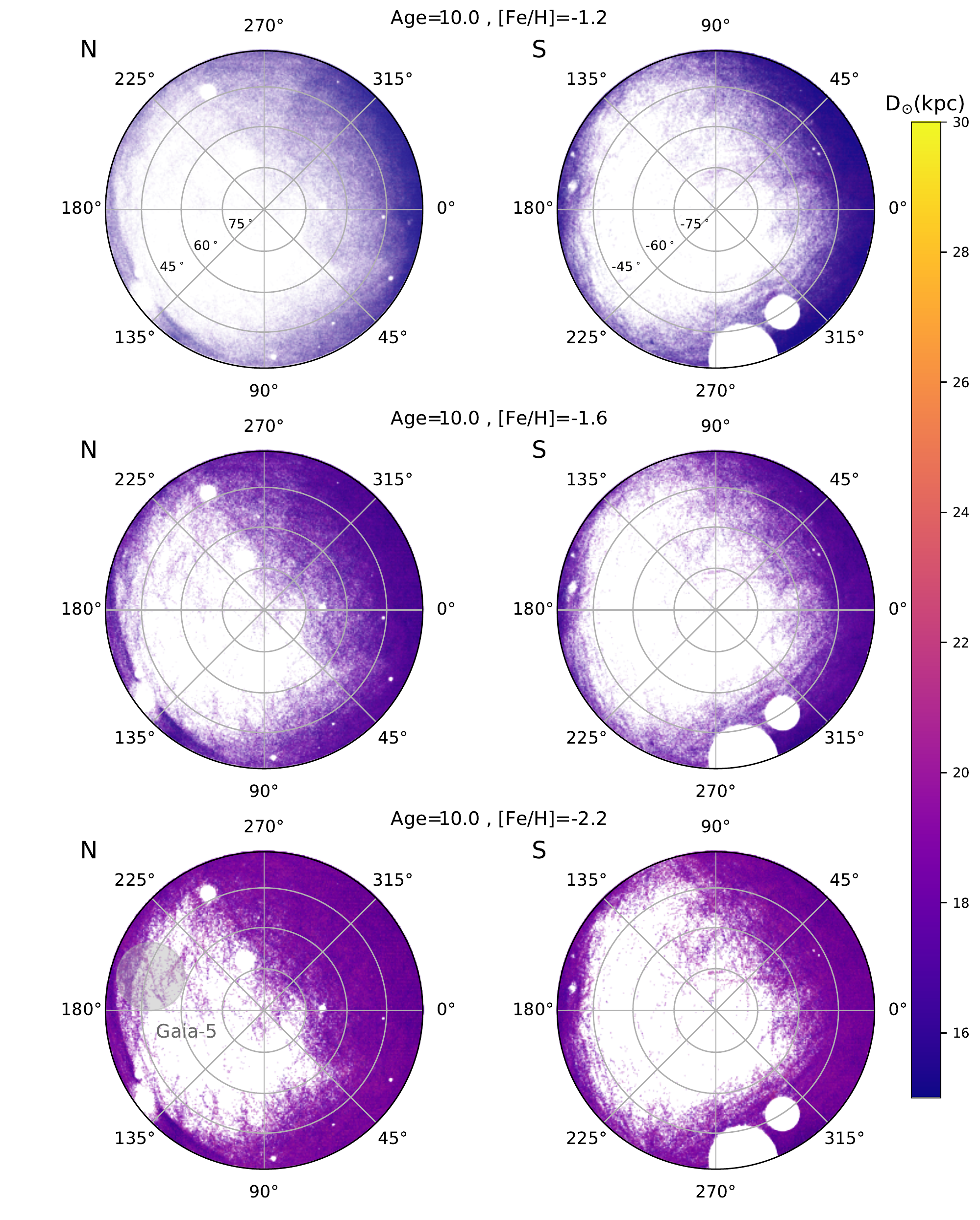}
\vspace{-1.0cm}
\end{center}
\caption{Spatial distribution of stream candidates at intermediate distances. Here we show the stellar stream density map as obtained from the \texttt{STREAMFINDER} based on 3 representative isochrone models. Each row corresponds to a particular isochrone model of age (in Gyr) and metallicity, as labelled. The left panels represent the North side of the ZEA projection system and the right panels represent the South. The colour scale is proportional to the heliocentric distances to the stellar members of the detected structures obtained as a by-product from the \texttt{STREAMFINDER} analysis. All streams displayed here have detection significance $>5\sigma$. New high confidence stream detections are marked on the map.}
\label{fig:ZEA_intermediate_halo}
\end{figure*}

\begin{figure*}
\begin{center}
\includegraphics[width=\hsize]{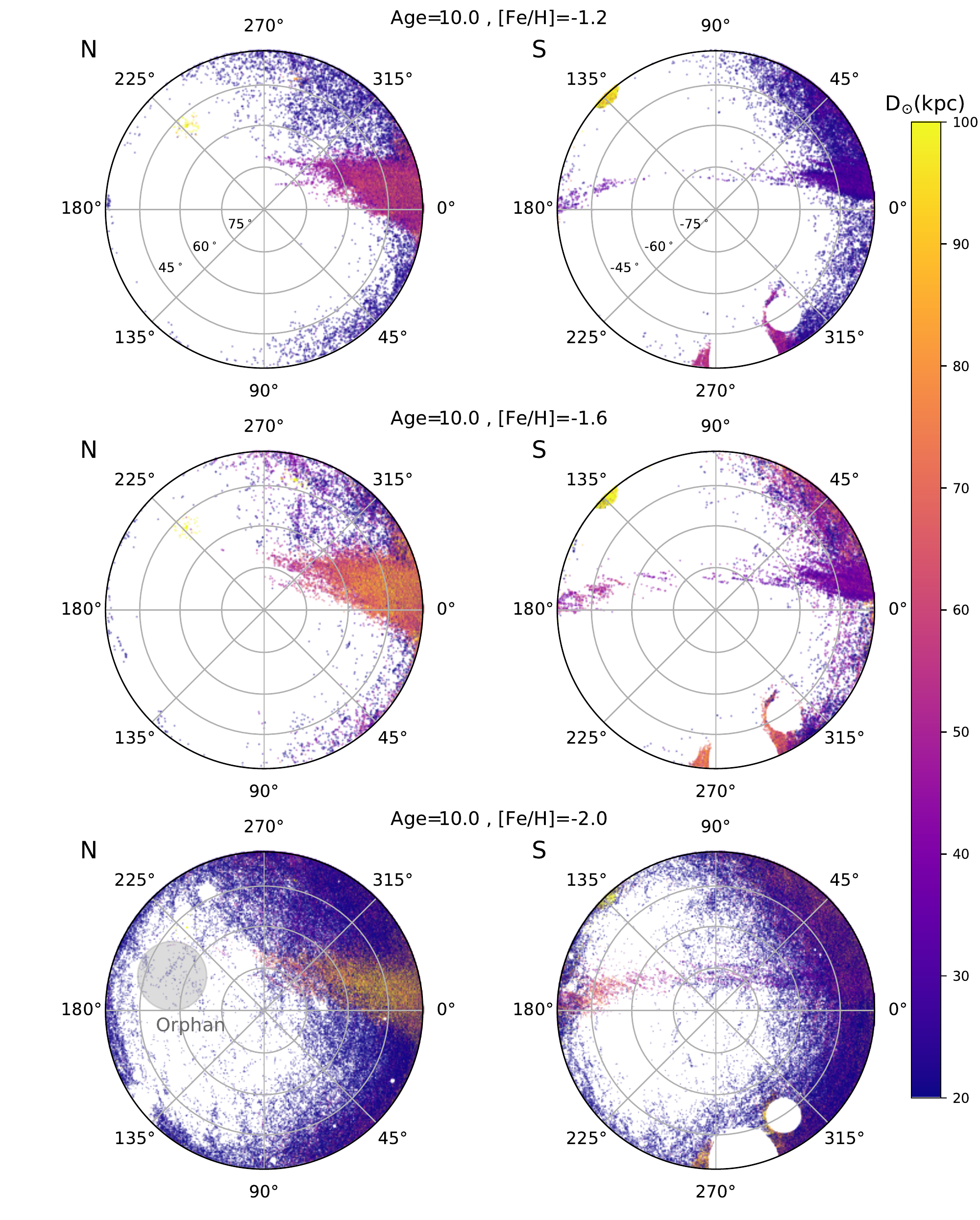}
\vspace{-1.0cm}
\end{center}
\caption{As Figure \ref{fig:ZEA_intermediate_halo}, but for the outer halo beyond $25\kpc$. The dominant structure seen out to large heliocentric distances in both hemispheres is the Sagittarius stream, which is detected despite the narrowness of the stream template model that we set in our algorithm. The interesting bifurcation of this structure is seen in the top-left panel. In addition, the lower-left panel shows an overdensity of stars in the region where the Orphan stream lies \citep{Grillmair2006Orphan}. Many other stream-like features are also detected, but most are confined to the nearer limit of the distance range shown.}
\label{fig:ZEA_outer_halo}
\end{figure*}

Figure~\ref{fig:ZEA_intermediate_halo} shows the results at intermediate distances in the halo in the range $D_{\odot}=[15,30]\kpc$ (again selecting ${\ln} \mathcal{L}_{\rm max}/{\ln} \mathcal{L}_{\eta=0} > 15$). Unlike Figure~\ref{fig:ZEA_inner_halo} that exhibits clearly distinguishable stream-like strings of stars, these maps produced at intermediate distances are rather fuzzy and only seldom show thin stream-like features. Some of these stream features become apparent in the regions $|b|>45\deg$ where the density of contaminating stars is low. The most obvious stream structure is Gaia-5, which is shaded in the grey circle in Figure~\ref{fig:ZEA_intermediate_halo} and its phase-space properties are presented in Figure \ref{fig:New_halo_candidates}.

The outer halo distribution, beyond $25\kpc$ is displayed in Figure~\ref{fig:ZEA_outer_halo} (again selecting ${\ln} \mathcal{L}_{\rm max}/{\ln} \mathcal{L}_{\eta=0} > 15$). The algorithm highlights a veritable deluge of stream-like structures, which are seen over a range of distances and metallicities. Comparison to Figure~\ref{fig:ZEA_Cecilia} shows that we detect the Sagittarius stream \citep{Ibata2001,Majewski2003SagStream} over a large swathe of the outer halo. This is somewhat surprising, since we set the stream model width to $100\pc$, which is appropriate for a globular cluster, but is actually a very poor template for this wide stream. We suspect that the spatial inhomogeneities in Gaia due its scanning law may partially explain the striated aspect of the Sagittarius stream in our maps (see, e.g., Figure~\ref{fig:ZEA_outer_halo}). The algorithm also detected a short arc of length $\sim10\deg$ of the $\sim60\deg$ long Orphan stream \citep{Grillmair2006Orphan} in our outer halo spatial maps (again, the chosen stream width of the model was not an appropriate template for this structure, which may explain why the full length was not recovered). For the position of the arc on the sky shown in Figure~\ref{fig:Known_streams}, we find the distance solutions for the Orphan stream members to be compatible with the study by \citet{Newberg2010OrphanFit}. Also, we find that its member stars have a tight proper motion distribution (Table \ref{tab:Stream_properties} provides proper motion values for the Orphan stream). This map also requires follow-up with radial velocity measurements in order to test the phase-space consistency of the other possible stream like structures that are distributed on these maps (for example, see the bottom panels of Figure \ref{fig:ZEA_outer_halo}).

\begin{figure*}
\begin{center}
\vbox{
\includegraphics[width=0.75\hsize]{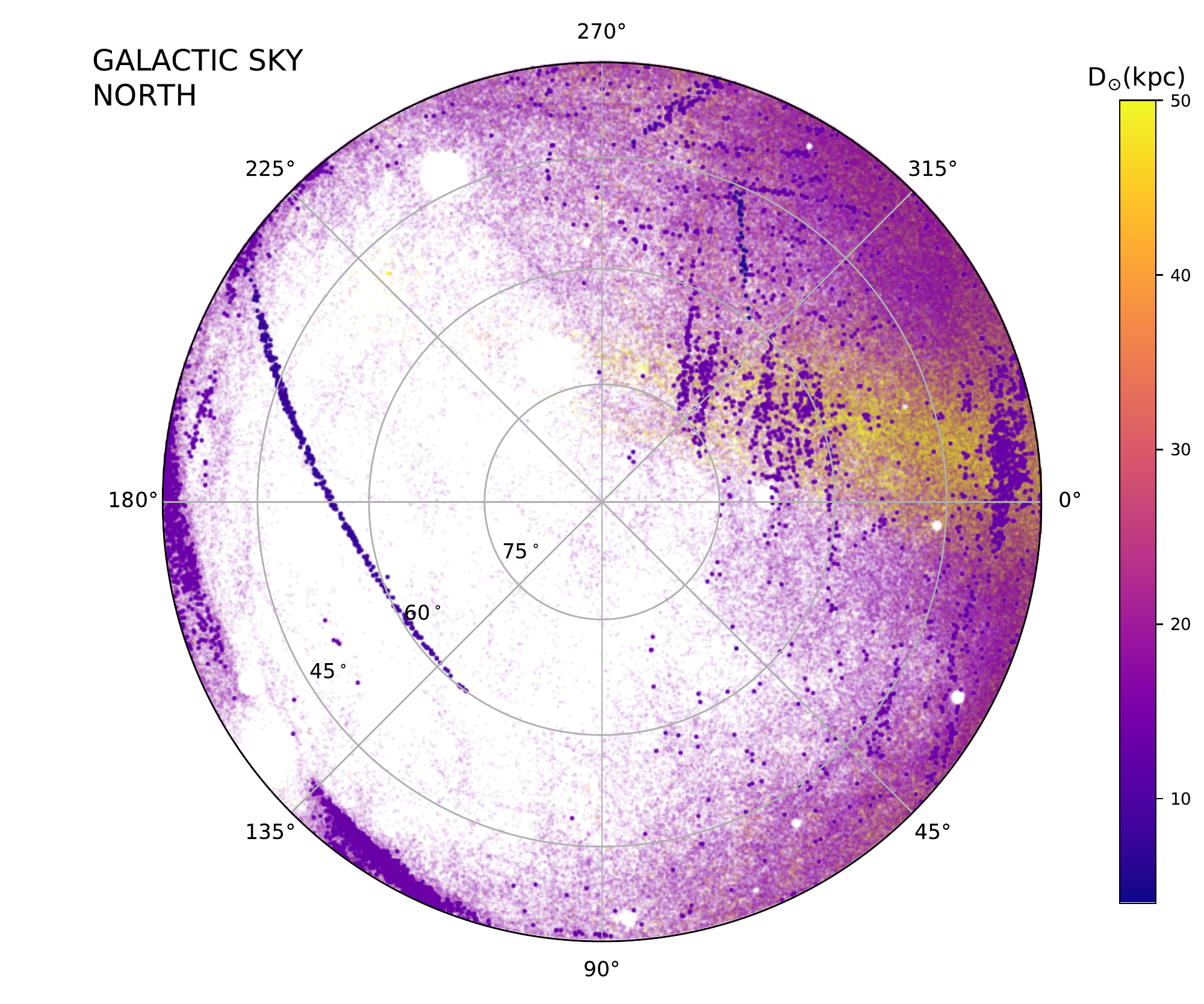}
\includegraphics[width=0.75\hsize]{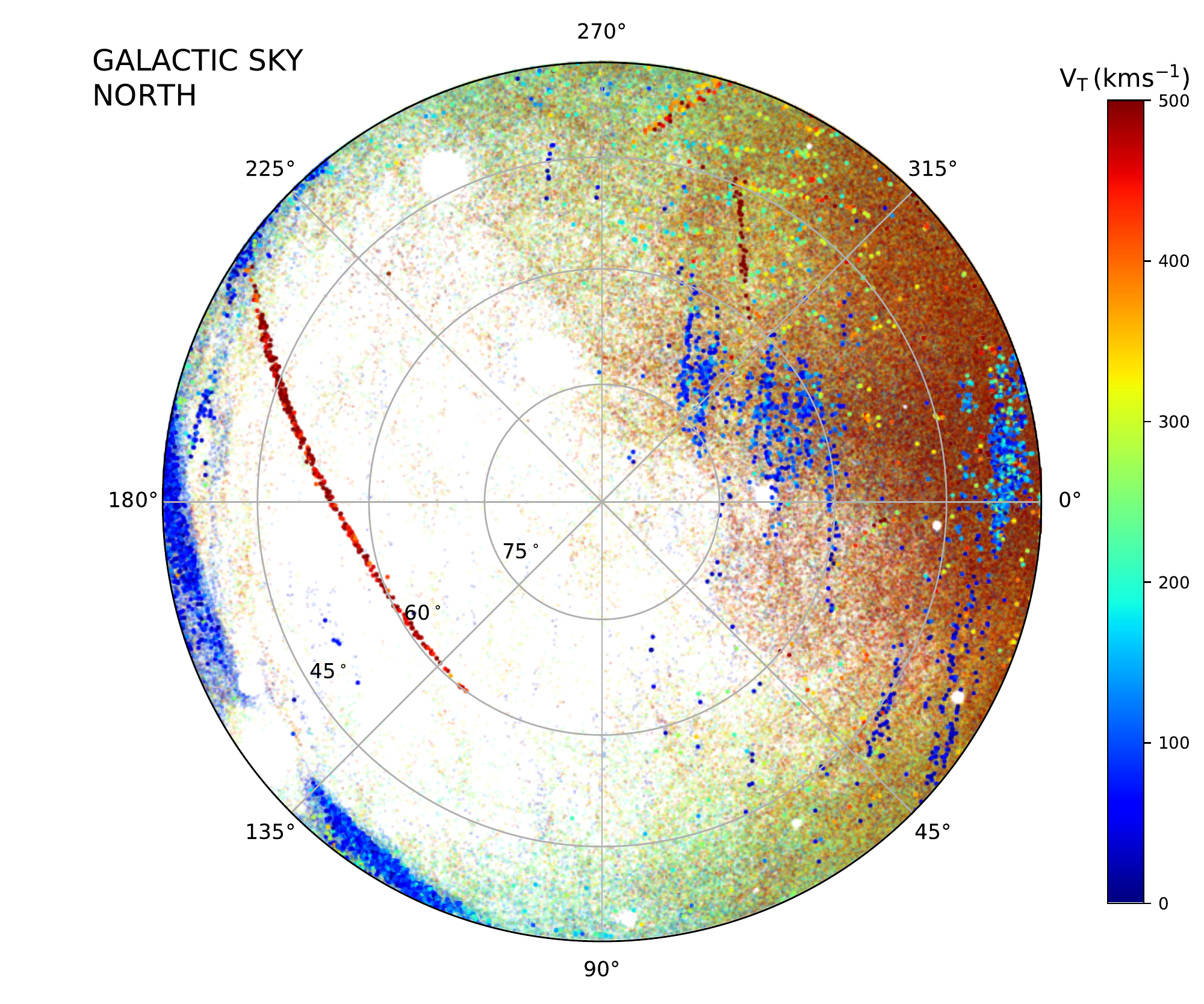}
}
\end{center}
\caption{Summary diagrams of the distance ($D_{\odot}>5\kpc$) and tangential velocity $V_T$ of stream-like structures in the northern Galactic sky. The tangential velocities are calculated based on the observed proper motion of the stars in DR2 and the corresponding distance estimates that we obtain from the algorithm. Most of the structures that we report here are visible in these diagrams, as are many others that we intend to investigate further in future contributions.} 
\label{fig:ZEA_North_Summary}
\end{figure*}

\begin{figure*}
\begin{center}
\vbox{
\includegraphics[width=0.75\hsize]{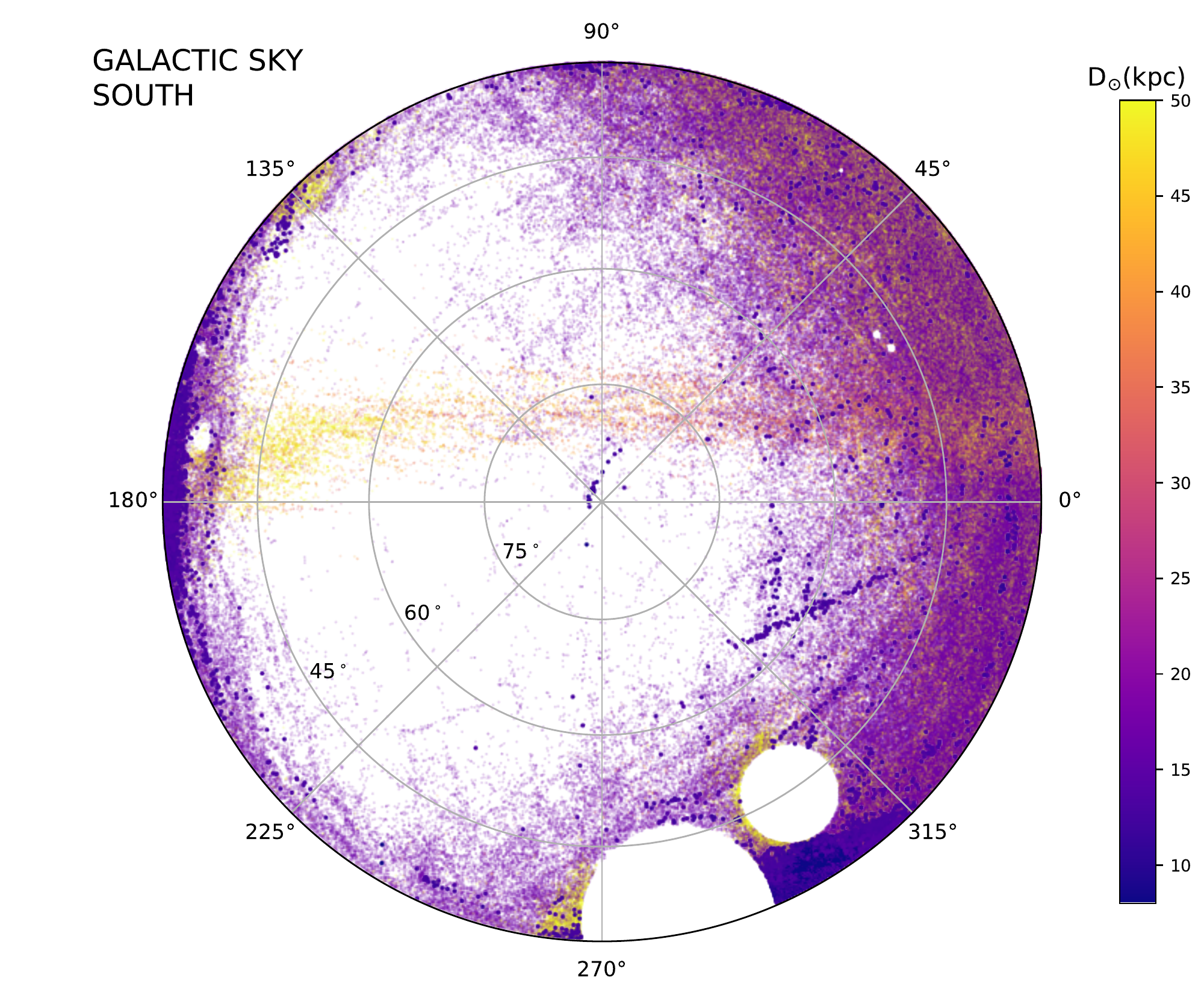}
\includegraphics[width=0.75\hsize]{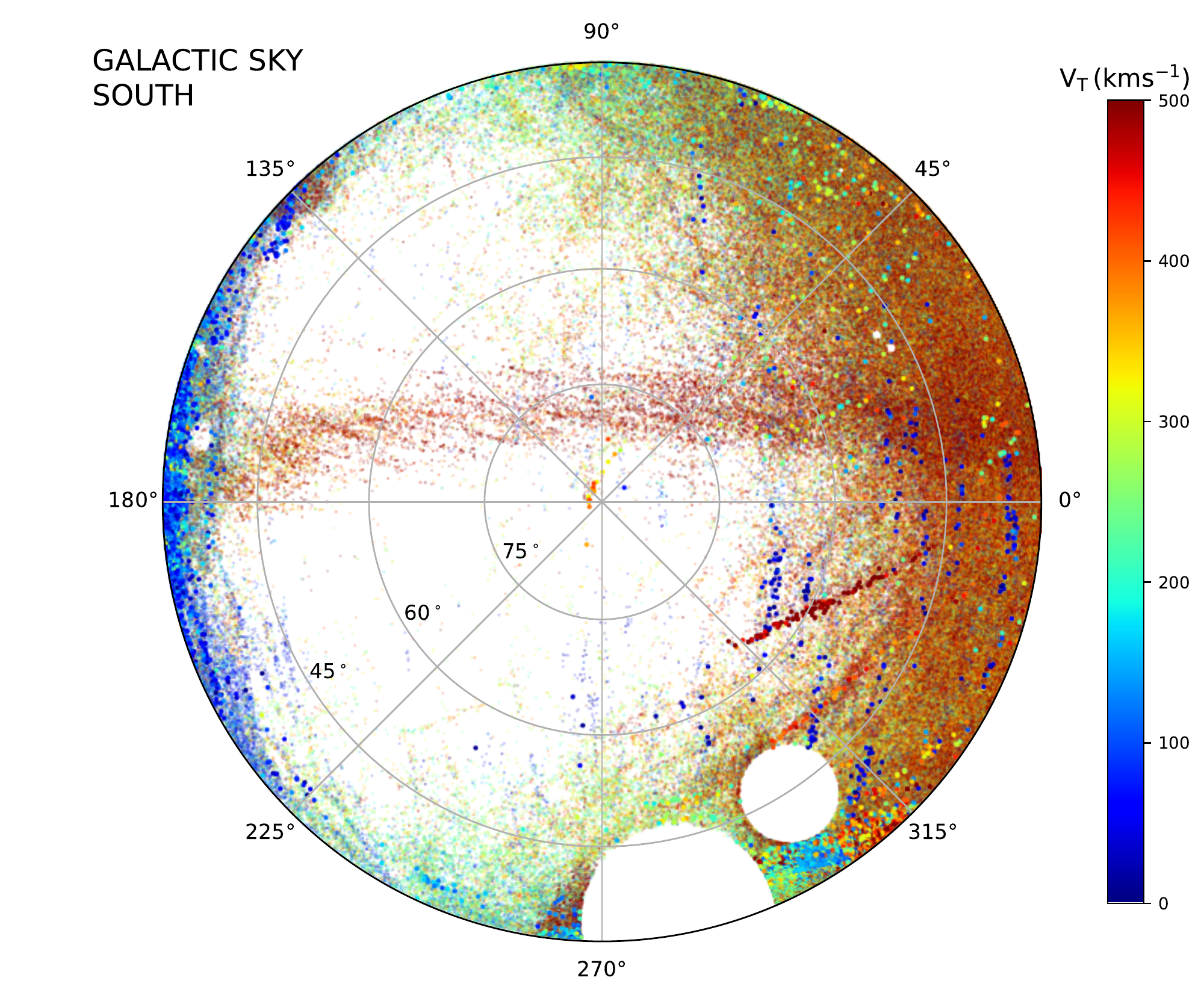}
}
\end{center}
\caption{As Figure~\ref{fig:ZEA_North_Summary}, but for the southern Galactic sky. }
\label{fig:ZEA_South_Summary}
\end{figure*}

Careful visual inspection of these maps indicated that the stream-like structures recovered by the algorithm are not associated with the extinction correction. In Figures~\ref{fig:ZEA_North_Summary} and \ref{fig:ZEA_South_Summary}, we present our summary plots made by combining the distance and metallicity samples for the north and south hemispheres, respectively. The top panels of these diagrams show the estimate of the distances of these structures (provided by the algorithm), while the bottom panels show an estimate of the magnitude of the tangential velocity calculated using the measured Gaia proper motions combined with the distance estimates. Many structures are beautifully resolved in this multi-parameter space.

Our aim in this contribution is not to present a thorough or complete census of halo streams (since it would require considerable more processing time to examine the necessary parameter space), but rather to present a preview of the large-scale stream structure of our Galaxy. Nevertheless, we have selected by hand a small number of structures that appear clearly in our maps, with kinematic properties that distinguish them from the contaminating Galactic population, and that are clearly not artefacts produced by Gaia's scanning law. A large number of other stream candidates have a clearly-defined stream-like morphology, but possess proper motions distributions that are similar to that of the halo, and we deem that they require further follow-up to be confident of their nature. 

The locations of the five structures we selected are marked in Figures~\ref{fig:ZEA_inner_halo} and \ref{fig:ZEA_intermediate_halo}, and their properties are shown in Figure~\ref{fig:New_halo_candidates}, and are also summarised in Table~\ref{tab:Stream_properties}. All these structures that we find have significance $>5\sigma$.

\subsection{Gaia-1}
Gaia-1 has an angular extent of $\sim15\deg$ and projected width of $\sim0.5\deg$. The orbital solutions provided by the algorithm imply that it is situated at a distance of $D_{\odot}\sim5.5\kpc$, which is in reasonable agreement with the Gaia parallax measurement of $0.216\pm0.038 \,\rm{mas}$ (i.e. $4.6\kpc$). This means that Gaia-1 has a physical width of $\sim40\pc$. The narrowness of the stream suggests that the progenitor likely is or was a globular cluster. Moreover, Gaia-1 has a strikingly high proper motion value of $\sim23.5\masyr$, implying that it has a transverse motion $\simgt 500\kms$. It will be worthwhile to measure the radial velocity of this system, as it may provide an interesting constraint on the Galactic potential simply from the requirement that the system is bound to the Milky Way.

\subsection{Gaia-2}
Gaia-2 turns out to be a considerably thin structure in our spatial maps. Extending over $\sim10\deg$ in length, we find that it possesses a distance gradient ranging from $D_{\odot}=[10$--$13]\kpc$. Given its narrowness and the location in the halo, we also suspect it to be a remnant of a globular cluster. We find Gaia-2 to be a highly coherent structure in proper motion space with an average proper motion magnitude of $\sim6.5\masyr$ and proper motion dispersion of $\sim0.75\masyr$.

\subsection{Gaia-3}
Gaia-3 can be easily identified as an isolated stream structure in Figure \ref{fig:ZEA_inner_halo}. In Figure~\ref{fig:New_halo_candidates} (third row), Gaia-3 clearly shows two distinct possible sets of distance solutions. The separation of these two different sets of solutions in position, distance and colour-magnitude distribution space, while not so much in proper motion space, suggests that what we detect here as Gaia-3 might in fact be a superposition of two streams, or a more complicated structure aligned along the line of sight. We shall describe this structure collectively here.

Gaia-3 is found to be extended over $\sim16\deg$ in sky with a distance range of $D_{\odot}=[9$--$14\kpc]$ with an average proper motion magnitude of $\sim 7.4\masyr$. Given its peculiarity, as suggested above and shown in Figure~\ref{fig:New_halo_candidates}, it is hard to comment on its physical width or the progenitor. The distance estimate of this structure too was found to be in good agreement with the Gaia parallax measurement of $0.101\pm0.013\,\rm{mas}$ (i.e. $9.8\kpc$).

\subsection{Gaia-4}
Gaia-4 appears to be a fine linear structure, found at a distance of $\sim D_{\odot}=11\kpc$. Given its narrowness and distance, we suspect the progenitor to be a globular cluster. Although we find Gaia-4 sitting within the range of halo field stars in proper motion space with an average value of $\sim 0.36\masyr$ (and proper motion dispersion of  $\sim0.70\masyr$), the fact that it emerges as a highly coherent structure in our maps makes it a confident structure. Here, we detect it as a very cold structure in proper motion space.

\subsection{Gaia-5}
\begin{figure*}
\begin{center}
\includegraphics[width=1.0\hsize]{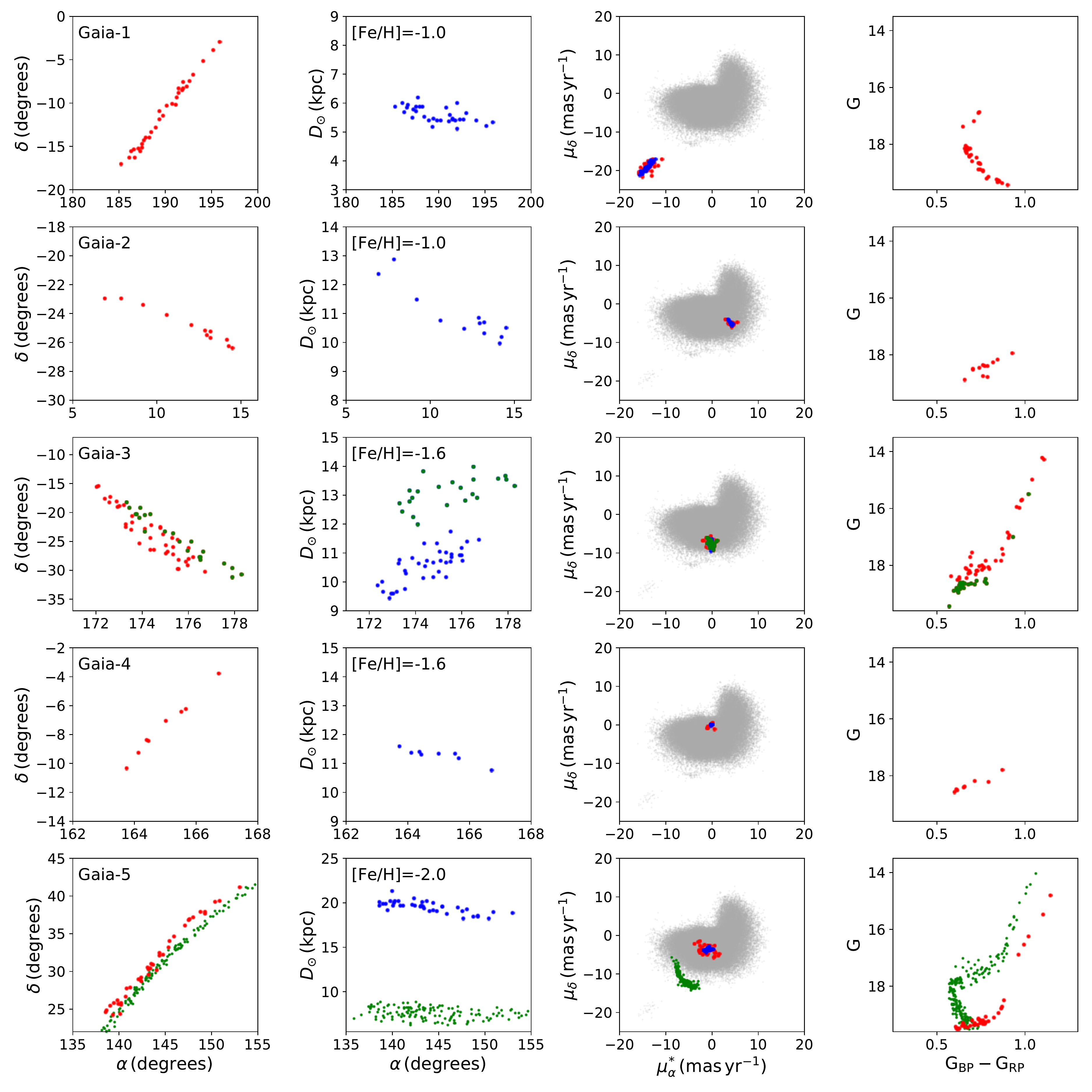}
\end{center}
\caption{As Figure \ref{fig:Known_streams} but for the selected set of newly-discovered streams. Oddly, for Gaia-3, we found two distinct possible sets of solutions based on distance estimates that we obtained from our algorithm, as highlighted in the respective panel. The more distant stars are coloured in green, while the relatively nearby ones are shown in red. This clear distinction of these two different sets of solutions in position, distance and colour-magnitude distribution space, while not so much in proper motion space, suggests that what we detect here as Gaia-3 might in fact be a superposition of two streams, or a more complicated structure aligned along the line of sight. The bottom row shows the properties of Gaia-5, which is found to lie parallel, but slightly offset, to GD-1 (shown on this bottom row in green). Nevertheless, it is very distinct from GD-1 both in its proper motion distribution and in its colour-magnitude distribution.}
\label{fig:New_halo_candidates}
\end{figure*}

We include Gaia-5 here as another interesting detection (bottom row panels in Figure~\ref{fig:New_halo_candidates}), as it is parallel to the GD-1 stream, and could easily have been confused with GD-1 without Gaia's excellent proper motion measurements. The properties of this object are shown in red for positions, observed proper motions and photometry, and in blue for distance and and proper motion orbital solutions. We also include the properties of GD-1 (in green) for comparison. The proper motions, along with the distance solutions, of Gaia-5 stars are distributed over a compact region that is very far from the region inhabited by GD-1; also the two colour-magnitude distributions (CMDs) are very different and well separated. Hence, unlike the possible bimodal stream distribution that we recognise in Gaia-3, we identify Gaia-5 as a stream unrelated to GD-1. The (error-weighted mean) parallax value we calculate for this structure would imply that it is substantially closer to the Sun than GD-1, which is both inconsistent with the model solutions of $\sim 20\kpc$, and is very difficult to reconcile with the CMD. However, our simple combination of the parallax measurements is highly susceptible to contaminants, which may explain the inconsistency.

We plan to examine these structures (and the many other stream candidates visible in Figures~\ref{fig:ZEA_North_Summary} and ~\ref{fig:ZEA_South_Summary}) in detail in later contributions. Careful analysis based on their astrometry and photometry, along with the mapping of these structures in  deeper astrophysical catalogues (e.g. SDSS, PS1, DES), would render a fuller insight into their origin, orbits and phase-space distribution. Some of the previously-known streams and new detections appear to present spatial  kinks, which is probably the effect of low-number statistics.

\begin{table*}
\caption{Parameters of the stellar streams. The ``Position'' column gives the extent of these structures, `$D_{\odot}$' is the approximate range of the distance solution as obtained by our algorithm, while column 4 lists the range of observed proper motion of the structure in the 2D proper motion space. The parallax $\pi$ is an uncertainty-weighted average of the Gaia measurements; for those objects where the parallax uncertainty is less than 33\% of the parallax, we also provide the corresponding distance. The discrepancy between the model distances and mean parallax measurement for the cases of Indus and Gaia-5 may be due to contaminants in the samples affecting the simple weighted average parallax reported here.}
\label{tab:Stream_properties}
\begin{center}
\begin{tabular}{lccccc}
\hline
\hline
Name & Position & $D_{\odot}$ (model) & ($\mu^*_{\alpha},\mu_{\delta}$) & $\pi$ & $\dfrac{1}{\pi}$\\
            & (extent) & ($\kpc$)   & ($\masyr$) & ($\rm{mas}$) & ($\kpc$)\\
\hline
GD-1 & $135\deg <\alpha < 190\deg$ & $6.5-10$ & $([-9.0, -3.0],[-14.0, -6.0])$ & $0.107\pm 0.010$  &  9.3  \\
    & $17\deg <\delta< 58\deg$ & &\\
    
Jhelum & $320\deg <\alpha < 360\deg$ & $11.7-15$ & $([5.0, 8.0],[-7.0 , -3.0])$ & $0.086\pm0.013$   & 11.6 \\
       & $-53\deg <\delta < -47\deg$& &  &\\
       
Indus & $320\deg <\alpha < 360\deg$ & $16-18$ & $([0.50, 6.0], [-8.0,-2.0])$ & $0.167\pm0.013$    & 6.0\\
     & $-67\deg <\delta < -53\deg$ & \\

Orphan & $145\deg <\alpha < 153\deg$ & $33-38$ & $([-1.0, -0.5], [-0.7,-0.1])$ & $-0.006\pm0.022$  &  \\
 				& $20\deg <\delta < 40\deg$\\

\hline

Gaia-1 & $184\deg <\alpha < 197\deg$   & $5 - 6$   &   $([-16.0, -11.0], [-22.0,-17.0])$ & $0.216\pm0.038$  &  4.6\\
       & $-18\deg <\delta < -2\deg$\\
       
Gaia-2 & $6\deg <\alpha < 15\deg$      & $10 - 13$   &  $([2.7, 5.4], [-6.0,-4.0])$ &  $0.117\pm0.062$ & \\
       & $-27\deg <\delta < -22\deg$\\
       
Gaia-3 & $171\deg <\alpha < 179\deg$   & $9 - 14$   &   $([-2.0, 1.0], [-9.3,-5.5])$ &   $0.101\pm 0.013$ & 9.9\\
       & $-32\deg <\delta < -15\deg$\\
       
Gaia-4& $163\deg <\alpha < 167\deg$  & $10.7-11.5$   &  $([-1.1, 0.5], [-1.1,0.6])$ &   $0.006\pm0.105$ & \\
       & $-11\deg <\delta < -3\deg$\\

Gaia-5& $137\deg <\alpha < 154\deg$ & $18.5-20.5$ &   $([-4.0,1.5], [-5.7,-1.5])$ & $0.156\pm0.031$  & $6.4$\\
				& $23\deg <\delta < 42\deg$ \\

\hline
\end{tabular}
\end{center}
\end{table*}

\begin{figure}
\begin{center}
\includegraphics[width=1.0\hsize]{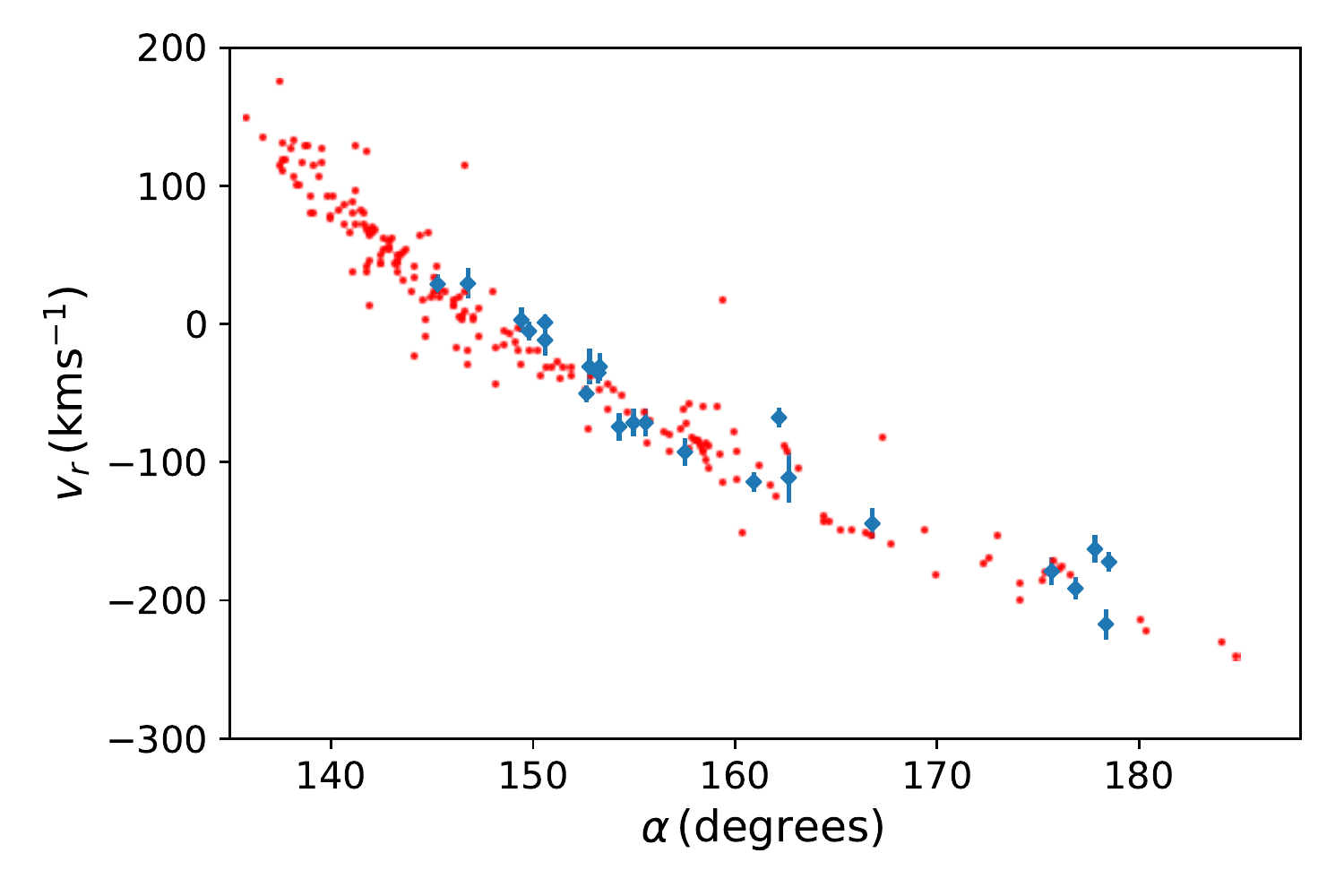}
\end{center}
\caption{Predicting the missing phase-space information of streams stars with \texttt{STREAMFINDER}. The red dots represent the radial velocity solutions of the GD-1 stars that are derived as a by-product of the application of the algorithm, whereas the blue markers are the observed radial velocities of GD-1 stars as tabulated by \citet{Koposov2010}. The \texttt{STREAMFINDER} sampled orbits in radial velocity space at intervals of $10\kms$ (which effectively causes an uncertainty of 10$\kms$ on the red dots). The good agreement with the observations illustrates the power of our algorithm in predicting the missing phase-space information of stream stars.}
\label{fig:GD1_vrad_comparison}
\end{figure}
\section{Discussions and Conclusions}
\label{sec:conclusions}

In this contribution, we present a new stellar stream map of the Milky Way halo, obtained by the application of our \texttt{STREAMFINDER} algorithm (described in Paper~I) on the recently published ESA/Gaia DR2 catalogue. This is the first time an all-sky structural and kinematic map of the stellar streams of the Milky Way halo has been constructed. Our algorithm detects numerous previously-known streams, which were discovered in much deeper photometric datasets (e.g. SDSS, PS1, DES), confirming that our method, which includes proper motion information, works as designed. Indeed, the fidelity of the GD-1 detection is striking, and reveals that the excellent Gaia proper motions provide very powerful discrimination. 

In addition, we find a large number of streams and stream candidates throughout the distance range probed. In this first exploration, we selected five good streams (named here as Gaia-1,2,3,4,5), to showcase the results, but many other candidates will require careful follow-up. In particular, the fact that Gaia scans the sky along great circles, but with an inhomogeneous number of visits, causes density inhomogeneities that appear like great circle streaks on the sky. This could cause some spurious stream detections (although the kinematics test in the \texttt{STREAMFINDER} algorithm should allow us to reject most such fake streams). Nevertheless, these spatial inhomogeneities in the Gaia DR2 necessarily make the survey noise properties very complex, invalidating the assumptions behind our ${\ln} \mathcal{L}_{\rm max}/{\ln} \mathcal{L}_{\eta=0} > 15$ selection criterion. This means, unfortunately, that the effective stream detection threshold is not uniform in our sky maps, and the significance of the detections is lower in regions where the Gaia inhomogeneities are more pronounced. 

A further caveat relates to the model distances we report. These distances are calculated by the algorithm based on an assumption of the metallicity of the stream stars. We expect that we do, in fact, have some ability to estimate  the metallicity of the candidate streams with our procedure, since using the correct metallicity model should enhance the contrast of the streams. This is borne out, for instance, for the case of GD-1, where we recover the largest number of stream stars when using the model corresponding to the actual metallicity of the system. Nevertheless, this is a poor substitute for actual metallicity measurements. Ongoing sky surveys, such as the Canada-France Imaging Survey (CFIS; \citealt{CFIS_2017_1,CFIS_2017_2}), or future large photometric surveys such as LSST \citep{LSST_Collab2012}, can help overcome this issue by providing good photometric metallicities that will break the distance degeneracy (and improve stream detection). The third Gaia data release (DR3), currently planned for 2020, will provide low-resolution prism spectra, also allowing metallicity measurements for the brighter stars.

As we showed in Paper~I, our algorithm naturally delivers the possible set of orbital solutions of the detected stream structures. This means that the algorithm predicts both the radial velocities and the distances of the stream stars. In Figure~\ref{fig:GD1_vrad_comparison} we use the orbital solutions to the GD-1 stream to demonstrate that this works very well: the predicted \texttt{STREAMFINDER} radial velocities match the stream velocities measured by \citet{Koposov2010}. Furthermore, our parallax measurement of $0.107\pm0.010\,\rm{mas}$ for GD-1, based on the sample we obtain with the \texttt{STREAMFINDER}, also matches well the distance range of the orbital solutions shown in Figure~\ref{fig:Known_streams} (these are not independent measurements, however the algorithm ``sees'' the potential stream stars diluted in a gigantic Galactic contaminating population). This success gives us confidence that we will be able to use the predicted \texttt{STREAMFINDER} radial velocities to probe the orbital properties of the stellar stream population as a whole.

Several more streams have been reported within $40\kpc$ than the five that we recover here (see Figure~\ref{fig:ZEA_Cecilia}). The reason for this is likely to be due, in part, to the specific parameter choices we adopted in the algorithm (for instance we chose a model width of $100\pc$ throughout, and we examined only a narrow range of stellar population template models). In subsequent contributions, we intend to relax these constraints allowing for a more complete census to be established. Additionally, we intend to examine different models of the Galactic potential; presumably our stream detection method should reveal the highest contrast for long stellar streams when using the correct potential. However, another reason that we did not recover all known streams within $40\kpc$ is simply that Gaia's photometry is not as deep as existing sky surveys; note that for a stellar population of metallicity ${\rm [Fe/H]=-1.5}$, the distance at which the proper motion uncertainties in Gaia DR2 at the main sequence turnoff are $50\kms$ (i.e. approximately half the dispersion of the contaminating halo) is $14.0 \kpc$. Hence it is not very surprising that photometric surveys that are much deeper than Gaia remain competitive for finding low-mass stellar streams at distances $\simgt 15 \kpc$.

Thanks to the amazingly rich phase-space information provided by the Gaia spacecraft and consortium, we are now starting to unravel the very fine details of galaxy formation in action. While the results presented here are but a first step in the comprehensive mapping of the Milky Way's stellar halo and accretion events, they already show the promises borne out by the deep, multi-dimensional space unveiled in DR2. The harvest of previously unknown thin stellar streams, likely stemming from the tidal disruption of globular clusters, opens up exciting times as these are powerful probes of the distribution of dark matter sub-halos in our surroundings \citep{Johnston2002subhalos, 2002MNRAS.332..915I, StreamGap_Carlberg2012,Bovy2016_DMsubhalo}, they can provide an independent inference of the location of the Sun in phase space \citep{Malhan2017}, and they can be used as sensitive seismographs to constrain the shape and depth of the Milky Way potential \citep{Ibata2013MW_halo_shape,Bonaca_Hogg2018}. The combination of Gaia DR2 and detections provided by \texttt{STREAMFINDER} places us in a unique position to disentangle the numerous detections accretion events in the Milky Way halo and open the most exciting Galactic archaeology playground to date.

\section*{ACKNOWLEDGEMENTS}

We thank the anonymous referee very much for their helpful comments.

This work has made use of data from the European Space Agency (ESA) mission {\it Gaia} (\url{https://www.cosmos.esa.int/gaia}), processed by the {\it Gaia} Data Processing and Analysis Consortium (DPAC, \url{https://www.cosmos.esa.int/web/gaia/dpac/consortium}). Funding for the DPAC has been provided by national institutions, in particular the institutions participating in the {\it Gaia} Multilateral Agreement. 

The authors would like to thank Michel Ringenbach of the HPC centre of the Universit\'e de Strasbourg for his kind support. We also acknowledge support by the Programme National Cosmology et Galaxies (PNCG) of CNRS/INSU with INP and IN2P3, co-funded by CEA and CNES. N. F. Martin gratefully acknowledges the Kavli Institute for Theoretical Physics in Santa Barbara and the organizers of the ``Cold Dark Matter 2018'' program, during which some of this work was performed. This research was supported in part by the National Science Foundation under Grant No. NSF PHY11-25915




\bibliographystyle{mnras}
\bibliography{ref1} 

\begin{thebibliography}{}
\makeatletter
\relax
\def\mn@urlcharsother{\let\do\@makeother \do\$\do\&\do\#\do\^\do\_\do\%\do\~}
\def\mn@doi{\begingroup\mn@urlcharsother \@ifnextchar [ {\mn@doi@}
  {\mn@doi@[]}}
\def\mn@doi@[#1]#2{\def\@tempa{#1}\ifx\@tempa\@empty \href
  {http://dx.doi.org/#2} {doi:#2}\else \href {http://dx.doi.org/#2} {#1}\fi
  \endgroup}
\def\mn@eprint#1#2{\mn@eprint@#1:#2::\@nil}
\def\mn@eprint@arXiv#1{\href {http://arxiv.org/abs/#1} {{\tt arXiv:#1}}}
\def\mn@eprint@dblp#1{\href {http://dblp.uni-trier.de/rec/bibtex/#1.xml}
  {dblp:#1}}
\def\mn@eprint@#1:#2:#3:#4\@nil{\def\@tempa {#1}\def\@tempb {#2}\def\@tempc
  {#3}\ifx \@tempc \@empty \let \@tempc \@tempb \let \@tempb \@tempa \fi \ifx
  \@tempb \@empty \def\@tempb {arXiv}\fi \@ifundefined
  {mn@eprint@\@tempb}{\@tempb:\@tempc}{\expandafter \expandafter \csname
  mn@eprint@\@tempb\endcsname \expandafter{\@tempc}}}

\bibitem[\protect\citeauthoryear{{Balbinot} \& {Gieles}}{{Balbinot} \&
  {Gieles}}{2018}]{Balbinot2018M}
{Balbinot} E.,  {Gieles} M.,  2018, \mn@doi [\mnras] {10.1093/mnras/stx2708},
  \href {http://adsabs.harvard.edu/abs/2018MNRAS.474.2479B} {474, 2479}

\bibitem[\protect\citeauthoryear{{Belokurov} et~al.,}{{Belokurov}
  et~al.}{2006}]{Belokurov2006}
{Belokurov} V.,  et~al., 2006, \mn@doi [\apjl] {10.1086/504797}, \href
  {http://adsabs.harvard.edu/abs/2006ApJ...642L.137B} {642, L137}

\bibitem[\protect\citeauthoryear{{Bernard} et~al.,}{{Bernard}
  et~al.}{2014}]{Bernard2014}
{Bernard} E.~J.,  et~al., 2014, \mn@doi [\mnras] {10.1093/mnrasl/slu089}, \href
  {http://adsabs.harvard.edu/abs/2014MNRAS.443L..84B} {443, L84}

\bibitem[\protect\citeauthoryear{{Bonaca} \& {Hogg}}{{Bonaca} \&
  {Hogg}}{2018}]{Bonaca_Hogg2018}
{Bonaca} A.,  {Hogg} D.~W.,  2018, preprint, \href
  {http://adsabs.harvard.edu/abs/2018arXiv180406854B} {} (\mn@eprint {arXiv}
  {1804.06854})

\bibitem[\protect\citeauthoryear{{Bovy}}{{Bovy}}{2016}]{Bovy2016_DMsubhalo}
{Bovy} J.,  2016, \mn@doi [Physical Review Letters]
  {10.1103/PhysRevLett.116.121301}, \href
  {http://adsabs.harvard.edu/abs/2016PhRvL.116l1301B} {116, 121301}

\bibitem[\protect\citeauthoryear{{Bovy}, {Bahmanyar}, {Fritz}  \&
  {Kallivayalil}}{{Bovy} et~al.}{2016}]{Bovy2016GD1Pal5}
{Bovy} J.,  {Bahmanyar} A.,  {Fritz} T.~K.,   {Kallivayalil} N.,  2016, \mn@doi
  [\apj] {10.3847/1538-4357/833/1/31}, \href
  {http://adsabs.harvard.edu/abs/2016ApJ...833...31B} {833, 31}

\bibitem[\protect\citeauthoryear{{Bullock} \& {Boylan-Kolchin}}{{Bullock} \&
  {Boylan-Kolchin}}{2017}]{Bullock2017problems}
{Bullock} J.~S.,  {Boylan-Kolchin} M.,  2017, \mn@doi [\araa]
  {10.1146/annurev-astro-091916-055313}, \href
  {http://adsabs.harvard.edu/abs/2017ARA%26A..55..343B} {55, 343}

\bibitem[\protect\citeauthoryear{{Carlberg}, {Grillmair}  \&
  {Hetherington}}{{Carlberg} et~al.}{2012}]{StreamGap_Carlberg2012}
{Carlberg} R.~G.,  {Grillmair} C.~J.,   {Hetherington} N.,  2012, \mn@doi
  [\apj] {10.1088/0004-637X/760/1/75}, \href
  {http://adsabs.harvard.edu/abs/2012ApJ...760...75C} {760, 75}

\bibitem[\protect\citeauthoryear{{Dehnen} \& {Binney}}{{Dehnen} \&
  {Binney}}{1998}]{Dehnen1998Massmodel}
{Dehnen} W.,  {Binney} J.,  1998, \mn@doi [\mnras]
  {10.1046/j.1365-8711.1998.01282.x}, \href
  {http://adsabs.harvard.edu/abs/1998MNRAS.294..429D} {294, 429}

\bibitem[\protect\citeauthoryear{{Erkal}, {Belokurov}, {Bovy}  \&
  {Sanders}}{{Erkal} et~al.}{2016}]{StreamGap_Erkal2016}
{Erkal} D.,  {Belokurov} V.,  {Bovy} J.,   {Sanders} J.~L.,  2016, \mn@doi
  [\mnras] {10.1093/mnras/stw1957}, \href
  {http://adsabs.harvard.edu/abs/2016MNRAS.463..102E} {463, 102}

\bibitem[\protect\citeauthoryear{{Evans, D. W.} et~al.,}{{Evans, D. W.}
  et~al.}{2018}]{GaiaDR2_2018_photometry}
{Evans, D. W.} et~al., 2018, \mn@doi [A\&A] {10.1051/0004-6361/201832756}

\bibitem[\protect\citeauthoryear{{Eyre} \& {Binney}}{{Eyre} \&
  {Binney}}{2009}]{EyreBinney2009}
{Eyre} A.,  {Binney} J.,  2009, \mn@doi [\mnras]
  {10.1111/j.1365-2966.2009.15494.x}, \href
  {http://adsabs.harvard.edu/abs/2009MNRAS.400..548E} {400, 548}

\bibitem[\protect\citeauthoryear{{Gaia Collaboration} et~al.,}{{Gaia
  Collaboration} et~al.}{2016}]{GaiaDR12016}
{Gaia Collaboration} et~al., 2016, \mn@doi [\aap]
  {10.1051/0004-6361/201629512}, \href
  {http://adsabs.harvard.edu/abs/2016A%26A...595A...2G} {595, A2}

\bibitem[\protect\citeauthoryear{{Gaia Collaboration}, {Brown, A. G. A.},
  {Vallenari, A.}, {Prusti, T.}, {de Bruijne, J. H. J.}  \& {et al.}}{{Gaia
  Collaboration} et~al.}{2018}]{GaiaDR2_2018_Brown}
{Gaia Collaboration} {Brown, A. G. A.} {Vallenari, A.} {Prusti, T.} {de
  Bruijne, J. H. J.}  {et al.} 2018, \mn@doi [A\&A]
  {10.1051/0004-6361/201833051}

\bibitem[\protect\citeauthoryear{{Grillmair}}{{Grillmair}}{2006}]{Grillmair2006Orphan}
{Grillmair} C.~J.,  2006, \mn@doi [\apjl] {10.1086/505863}, \href
  {http://adsabs.harvard.edu/abs/2006ApJ...645L..37G} {645, L37}

\bibitem[\protect\citeauthoryear{{Grillmair} \& {Carlin}}{{Grillmair} \&
  {Carlin}}{2016}]{Grillmair2016BookChapter}
{Grillmair} C.~J.,  {Carlin} J.~L.,  2016, in {Newberg} H.~J.,  {Carlin} J.~L.,
   eds,  Astrophysics and Space Science Library Vol. 420, Tidal Streams in the
  Local Group and Beyond. p.~87 (\mn@eprint {arXiv} {1603.08936}),
  \mn@doi{10.1007/978-3-319-19336-6_4}

\bibitem[\protect\citeauthoryear{{Grillmair} \& {Dionatos}}{{Grillmair} \&
  {Dionatos}}{2006}]{Grillmair2006GD1_correct}
{Grillmair} C.~J.,  {Dionatos} O.,  2006, \mn@doi [\apjl] {10.1086/505111},
  \href {http://adsabs.harvard.edu/abs/2006ApJ...643L..17G} {643, L17}

\bibitem[\protect\citeauthoryear{{Harris}}{{Harris}}{2010}]{Harris2010_MW_GC}
{Harris} W.~E.,  2010, preprint, \href
  {http://adsabs.harvard.edu/abs/2010arXiv1012.3224H} {} (\mn@eprint {arXiv}
  {1012.3224})

\bibitem[\protect\citeauthoryear{{Helmi} \& {White}}{{Helmi} \&
  {White}}{1999}]{Helmi1999BuildingHalo}
{Helmi} A.,  {White} S.~D.~M.,  1999, in {Gibson} B.~K.,  {Axelrod} R.~S.,
  {Putman} M.~E.,  eds,  Astronomical Society of the Pacific Conference Series
  Vol. 165, The Third Stromlo Symposium: The Galactic Halo. p.~89 (\mn@eprint
  {} {astro-ph/9811108})

\bibitem[\protect\citeauthoryear{{Helmi, A.}, {van Leeuwen, F.}, {McMillan, P.}
   \& {DPAC}}{{Helmi, A.} et~al.}{2018}]{GaiaDR2_2018_Pmotions}
{Helmi, A.} {van Leeuwen, F.} {McMillan, P.}  {DPAC} 2018, \mn@doi [A\&A]
  {10.1051/0004-6361/201832698}

\bibitem[\protect\citeauthoryear{{Ibata}, {Lewis}, {Irwin}, {Totten}  \&
  {Quinn}}{{Ibata} et~al.}{2001}]{Ibata2001}
{Ibata} R.,  {Lewis} G.~F.,  {Irwin} M.,  {Totten} E.,   {Quinn} T.,  2001,
  \mn@doi [\apj] {10.1086/320060}, \href
  {http://adsabs.harvard.edu/abs/2001ApJ...551..294I} {551, 294}

\bibitem[\protect\citeauthoryear{Ibata, Lewis, Irwin  \& Quinn}{Ibata
  et~al.}{2002}]{2002MNRAS.332..915I}
Ibata R.~A.,  Lewis G.~F.,  Irwin M.~J.,   Quinn T.,  2002, Monthly Notices of
  the Royal Astronomical Society, 332, 915

\bibitem[\protect\citeauthoryear{{Ibata}, {Lewis}, {Martin}, {Bellazzini}  \&
  {Correnti}}{{Ibata} et~al.}{2013}]{Ibata2013MW_halo_shape}
{Ibata} R.,  {Lewis} G.~F.,  {Martin} N.~F.,  {Bellazzini} M.,   {Correnti} M.,
   2013, \mn@doi [\apjl] {10.1088/2041-8205/765/1/L15}, \href
  {http://adsabs.harvard.edu/abs/2013ApJ...765L..15I} {765, L15}

\bibitem[\protect\citeauthoryear{{Ibata} et~al.,}{{Ibata}
  et~al.}{2017a}]{CFIS_2017_1}
{Ibata} R.~A.,  et~al., 2017a, \mn@doi [\apj] {10.3847/1538-4357/aa855c}, \href
  {http://adsabs.harvard.edu/abs/2017ApJ...848..128I} {848, 128}

\bibitem[\protect\citeauthoryear{{Ibata} et~al.,}{{Ibata}
  et~al.}{2017b}]{CFIS_2017_2}
{Ibata} R.~A.,  et~al., 2017b, \mn@doi [\apj] {10.3847/1538-4357/aa8562}, \href
  {http://adsabs.harvard.edu/abs/2017ApJ...848..129I} {848, 129}

\bibitem[\protect\citeauthoryear{{Ibata}, {Malhan}, {Martin}  \&
  {Starkenburg}}{{Ibata} et~al.}{2018}]{Ibata_2018_Phlegethon}
{Ibata} R.~A.,  {Malhan} K.,  {Martin} N.~F.,   {Starkenburg} E.,  2018,
  preprint, \href {http://adsabs.harvard.edu/abs/2018arXiv180601195I} {}
  (\mn@eprint {arXiv} {1806.01195})

\bibitem[\protect\citeauthoryear{{Johnston}, {Hernquist}  \&
  {Bolte}}{{Johnston} et~al.}{1996}]{Johnston1996}
{Johnston} K.~V.,  {Hernquist} L.,   {Bolte} M.,  1996, \mn@doi [\apj]
  {10.1086/177418}, \href {http://adsabs.harvard.edu/abs/1996ApJ...465..278J}
  {465, 278}

\bibitem[\protect\citeauthoryear{{Johnston}, {Zhao}, {Spergel}  \&
  {Hernquist}}{{Johnston} et~al.}{1999}]{Johnston1999GalPot}
{Johnston} K.~V.,  {Zhao} H.,  {Spergel} D.~N.,   {Hernquist} L.,  1999,
  \mn@doi [\apjl] {10.1086/311876}, \href
  {http://adsabs.harvard.edu/abs/1999ApJ...512L.109J} {512, L109}

\bibitem[\protect\citeauthoryear{{Johnston}, {Spergel}  \& {Haydn}}{{Johnston}
  et~al.}{2002}]{Johnston2002subhalos}
{Johnston} K.~V.,  {Spergel} D.~N.,   {Haydn} C.,  2002, \mn@doi [\apj]
  {10.1086/339791}, \href {http://adsabs.harvard.edu/abs/2002ApJ...570..656J}
  {570, 656}

\bibitem[\protect\citeauthoryear{{Karim} \& {Mamajek}}{{Karim} \&
  {Mamajek}}{2017}]{R_sun_value}
{Karim} M.~T.,  {Mamajek} E.~E.,  2017, \mn@doi [\mnras]
  {10.1093/mnras/stw2772}, \href
  {http://adsabs.harvard.edu/abs/2017MNRAS.465..472K} {465, 472}

\bibitem[\protect\citeauthoryear{{Koposov}, {Rix}  \& {Hogg}}{{Koposov}
  et~al.}{2010}]{Koposov2010}
{Koposov} S.~E.,  {Rix} H.-W.,   {Hogg} D.~W.,  2010, \mn@doi [\apj]
  {10.1088/0004-637X/712/1/260}, \href
  {http://adsabs.harvard.edu/abs/2010ApJ...712..260K} {712, 260}

\bibitem[\protect\citeauthoryear{{K{\"u}pper}, {Balbinot}, {Bonaca},
  {Johnston}, {Hogg}, {Kroupa}  \& {Santiago}}{{K{\"u}pper}
  et~al.}{2015}]{Kupper2015}
{K{\"u}pper} A.~H.~W.,  {Balbinot} E.,  {Bonaca} A.,  {Johnston} K.~V.,  {Hogg}
  D.~W.,  {Kroupa} P.,   {Santiago} B.~X.,  2015, \mn@doi [\apj]
  {10.1088/0004-637X/803/2/80}, \href
  {http://adsabs.harvard.edu/abs/2015ApJ...803...80K} {803, 80}

\bibitem[\protect\citeauthoryear{{LSST Dark Energy Science
  Collaboration}}{{LSST Dark Energy Science
  Collaboration}}{2012}]{LSST_Collab2012}
{LSST Dark Energy Science Collaboration} 2012, preprint, \href
  {http://adsabs.harvard.edu/abs/2012arXiv1211.0310L} {} (\mn@eprint {arXiv}
  {1211.0310})

\bibitem[\protect\citeauthoryear{{Law} \& {Majewski}}{{Law} \&
  {Majewski}}{2010}]{LawMajewski2010}
{Law} D.~R.,  {Majewski} S.~R.,  2010, \mn@doi [\apj]
  {10.1088/0004-637X/714/1/229}, \href
  {http://adsabs.harvard.edu/abs/2010ApJ...714..229L} {714, 229}

\bibitem[\protect\citeauthoryear{{Lindegren, L.}, {Hernandez, J.}, {Bombrun,
  A.}, {Klioner, S.}, {Bastian, U.}  \& {Ramos-Lerate, M.}}{{Lindegren, L.}
  et~al.}{2018}]{GaiaDR2_2018_astrometry}
{Lindegren, L.} {Hernandez, J.} {Bombrun, A.} {Klioner, S.} {Bastian, U.}
  {Ramos-Lerate, M.} 2018, \mn@doi [A\&A] {10.1051/0004-6361/201832727}

\bibitem[\protect\citeauthoryear{{Luri, Xavier} et~al.,}{{Luri, Xavier}
  et~al.}{2018}]{GaiaDR2_2018_Parallaxes}
{Luri, Xavier} et~al., 2018, \mn@doi [A\&A] {10.1051/0004-6361/201832964}

\bibitem[\protect\citeauthoryear{{Majewski}, {Skrutskie}, {Weinberg}  \&
  {Ostheimer}}{{Majewski} et~al.}{2003}]{Majewski2003SagStream}
{Majewski} S.~R.,  {Skrutskie} M.~F.,  {Weinberg} M.~D.,   {Ostheimer} J.~C.,
  2003, \mn@doi [\apj] {10.1086/379504}, \href
  {http://adsabs.harvard.edu/abs/2003ApJ...599.1082M} {599, 1082}

\bibitem[\protect\citeauthoryear{{Malhan} \& {Ibata}}{{Malhan} \&
  {Ibata}}{2017}]{Malhan2017}
{Malhan} K.,  {Ibata} R.~A.,  2017, \mn@doi [\mnras] {10.1093/mnras/stx1618},
  \href {http://adsabs.harvard.edu/abs/2017MNRAS.471.1005M} {471, 1005}

\bibitem[\protect\citeauthoryear{{Malhan} \& {Ibata}}{{Malhan} \&
  {Ibata}}{2018}]{Malhan2018_SFI}
{Malhan} K.,  {Ibata} R.~A.,  2018, \mn@doi [\mnras] {10.1093/mnras/sty912},
  \href {http://adsabs.harvard.edu/abs/2018MNRAS.tmp..876M} {}

\bibitem[\protect\citeauthoryear{{Marigo}, {Girardi}, {Bressan}, {Groenewegen},
  {Silva}  \& {Granato}}{{Marigo} et~al.}{2008}]{Marigo2008Padova}
{Marigo} P.,  {Girardi} L.,  {Bressan} A.,  {Groenewegen} M.~A.~T.,  {Silva}
  L.,   {Granato} G.~L.,  2008, \mn@doi [\aap] {10.1051/0004-6361:20078467},
  \href {http://adsabs.harvard.edu/abs/2008A%26A...482..883M} {482, 883}

\bibitem[\protect\citeauthoryear{{Mateu}, {Read}  \& {Kawata}}{{Mateu}
  et~al.}{2018}]{Mateu2017}
{Mateu} C.,  {Read} J.~I.,   {Kawata} D.,  2018, \mn@doi [\mnras]
  {10.1093/mnras/stx2937}, \href
  {http://adsabs.harvard.edu/abs/2018MNRAS.474.4112M} {474, 4112}

\bibitem[\protect\citeauthoryear{{McConnachie}}{{McConnachie}}{2012}]{McConnachie2012}
{McConnachie} A.~W.,  2012, \mn@doi [\aj] {10.1088/0004-6256/144/1/4}, \href
  {http://adsabs.harvard.edu/abs/2012AJ....144....4M} {144, 4}

\bibitem[\protect\citeauthoryear{{Newberg}, {Willett}, {Yanny}  \&
  {Xu}}{{Newberg} et~al.}{2010}]{Newberg2010OrphanFit}
{Newberg} H.~J.,  {Willett} B.~A.,  {Yanny} B.,   {Xu} Y.,  2010, \mn@doi
  [\apj] {10.1088/0004-637X/711/1/32}, \href
  {http://adsabs.harvard.edu/abs/2010ApJ...711...32N} {711, 32}

\bibitem[\protect\citeauthoryear{{Reid} et~al.,}{{Reid}
  et~al.}{2014}]{Reid_Sun_circ_vel}
{Reid} M.~J.,  et~al., 2014, \mn@doi [\apj] {10.1088/0004-637X/783/2/130},
  \href {http://adsabs.harvard.edu/abs/2014ApJ...783..130R} {783, 130}

\bibitem[\protect\citeauthoryear{{Robin} et~al.,}{{Robin}
  et~al.}{2012}]{GUMS2012}
{Robin} A.~C.,  et~al., 2012, \mn@doi [\aap] {10.1051/0004-6361/201118646},
  \href {http://adsabs.harvard.edu/abs/2012A%26A...543A.100R} {543, A100}

\bibitem[\protect\citeauthoryear{{Sanders}, {Bovy}  \& {Erkal}}{{Sanders}
  et~al.}{2016}]{StreamGap_Sanders2016}
{Sanders} J.~L.,  {Bovy} J.,   {Erkal} D.,  2016, \mn@doi [\mnras]
  {10.1093/mnras/stw232}, \href
  {http://adsabs.harvard.edu/abs/2016MNRAS.457.3817S} {457, 3817}

\bibitem[\protect\citeauthoryear{Sanderson \& Curtin}{Sanderson \&
  Curtin}{2017}]{gmm2017armadillo}
Sanderson C.,  Curtin R.,  2017, Journal of Open Source Software, 2

\bibitem[\protect\citeauthoryear{{Schlegel}, {Finkbeiner}  \&
  {Davis}}{{Schlegel} et~al.}{1998}]{Schlegel_Extinc_maps1998}
{Schlegel} D.~J.,  {Finkbeiner} D.~P.,   {Davis} M.,  1998, \mn@doi [\apj]
  {10.1086/305772}, \href {http://adsabs.harvard.edu/abs/1998ApJ...500..525S}
  {500, 525}

\bibitem[\protect\citeauthoryear{{Sch{\"o}nrich}, {Binney}  \&
  {Dehnen}}{{Sch{\"o}nrich} et~al.}{2010}]{Schornich2010_Sun}
{Sch{\"o}nrich} R.,  {Binney} J.,   {Dehnen} W.,  2010, \mn@doi [\mnras]
  {10.1111/j.1365-2966.2010.16253.x}, \href
  {http://adsabs.harvard.edu/abs/2010MNRAS.403.1829S} {403, 1829}

\bibitem[\protect\citeauthoryear{{Shipp} et~al.,}{{Shipp}
  et~al.}{2018}]{DES_11_streams2018}
{Shipp} N.,  et~al., 2018, preprint, \href
  {http://adsabs.harvard.edu/abs/2018arXiv180103097S} {} (\mn@eprint {arXiv}
  {1801.03097})

\bibitem[\protect\citeauthoryear{{de Boer}, {Belokurov}, {Koposov},
  {Ferrarese}, {Erkal}, {Cote}  \& {Navarro}}{{de Boer}
  et~al.}{2018}]{Boer2018}
{de Boer} T.~J.~L.,  {Belokurov} V.,  {Koposov} S.~E.,  {Ferrarese} L.,
  {Erkal} D.,  {Cote} P.,   {Navarro} J.~F.,  2018, preprint, \href
  {http://adsabs.harvard.edu/abs/2018arXiv180108948D} {} (\mn@eprint {arXiv}
  {1801.08948})

\bibitem[\protect\citeauthoryear{{de Bruijne}}{{de
  Bruijne}}{2012}]{Gaia2012deBruijne}
{de Bruijne} J.~H.~J.,  2012, \mn@doi [\apss] {10.1007/s10509-012-1019-4},
  \href {http://adsabs.harvard.edu/abs/2012Ap%26SS.341...31D} {341, 31}

\makeatother
\end{thebibliography}


\bsp	
\label{lastpage}
\end{document}